\documentclass[sigconf]{acmart}

\AtBeginDocument{%
  \providecommand\BibTeX{{%
    \normalfont B\kern-0.5em{\scshape i\kern-0.25em b}\kern-0.8em\TeX}}}

\copyrightyear{2026}
\acmYear{2026}
\setcopyright{cc}
\setcctype{by}

\acmConference[DIS '26]{Designing Interactive Systems Conference}{June 13--17, 2026}{Singapore, Singapore}
\acmBooktitle{Designing Interactive Systems Conference (DIS '26), June 13--17, 2026, Singapore, Singapore}
\acmDOI{10.1145/3800645.3812903}
\acmISBN{979-8-4007-2563-0/2026/06}

\usepackage[utf8]{inputenc}
\usepackage{array}
\usepackage{booktabs}
\usepackage{enumitem}
\usepackage{tabularx}

\makeatletter
\patchcmd{\@specialsection}{\section*{#1}}{\vspace{-0.45\baselineskip}\section*{#1}}{}{}
\makeatother

\begin{document}

\title[Profy: Visualizing Expert--Amateur Motor Skill Differences]{Interpretable Visualization of Expertise-Dependent Motor Skills Toward Supporting Piano Practice}
\author{Kazuki Kawamura}
\affiliation{%
  \institution{The University of Tokyo}
  \city{Tokyo}
  \country{Japan}}
\affiliation{%
  \institution{Sony Computer Science Laboratories}
  \city{Kyoto}
  \country{Japan}}
\email{kwmr@acm.org}

\author{Fujiki Nakamura}
\affiliation{%
  \institution{The University of Tokyo}
  \city{Tokyo}
  \country{Japan}}
\email{fujiki-nakamura@g.ecc.u-tokyo.ac.jp}

\author{Hayato Nishioka}
\affiliation{%
  \institution{Sony Computer Science Laboratories}
  \city{Tokyo}
  \country{Japan}}
\affiliation{%
  \institution{NeuroPiano Institute}
  \city{Kyoto}
  \country{Japan}}
\email{nishioka@csl.sony.co.jp}

\author{Momoko Shioki}
\affiliation{%
  \institution{Sony Computer Science Laboratories}
  \city{Tokyo}
  \country{Japan}}
\affiliation{%
  \institution{NeuroPiano Institute}
  \city{Kyoto}
  \country{Japan}}
\email{mkpfmsp.9125@gmail.com}

\author{Shinichi Furuya}
\affiliation{%
  \institution{Sony Computer Science Laboratories}
  \city{Tokyo}
  \country{Japan}}
\affiliation{%
  \institution{NeuroPiano Institute}
  \city{Kyoto}
  \country{Japan}}
\email{furuya@csl.sony.co.jp}

\author{Jun Rekimoto}
\affiliation{%
  \institution{The University of Tokyo}
  \city{Tokyo}
  \country{Japan}}
\affiliation{%
  \institution{Sony Computer Science Laboratories}
  \city{Kyoto}
  \country{Japan}}
\email{rekimoto@acm.org}

\renewcommand{\shortauthors}{Kawamura et al.}

\begin{abstract}
The quality of piano performance depends on nuanced timing, articulation, and dynamic control, but practice feedback is often summary-based and hard to act on. We introduce Profy, a weakly supervised system that learns from take-level labels derived from aggregated listener ratings (expert-labeled vs.\ amateur-labeled) to produce time-aligned highlights for review during piano practice. We collected synchronized 1\,kHz key-motion and audio from 73 pianists and used 1{,}083 valid takes for modeling and evaluation. The model outputs clip-level predictions together with evidence scores on a shared resampled model time base for visualization. On 20 amateur clips from short technique studies annotated by 21 expert pianists, the displayed highlight score aligns with passages that expert pianists marked for review despite training without localized labels (Pearson $r{=}0.61$, ROC-AUC $0.75$). Rather than summarizing a take with a single global score, Profy helps learners decide where to inspect next by supporting scrubbing, looping, and focused replay of time-localized passages associated with expert--amateur differences.
\end{abstract}



\begin{teaserfigure}
\centering
  \includegraphics[width=\textwidth]{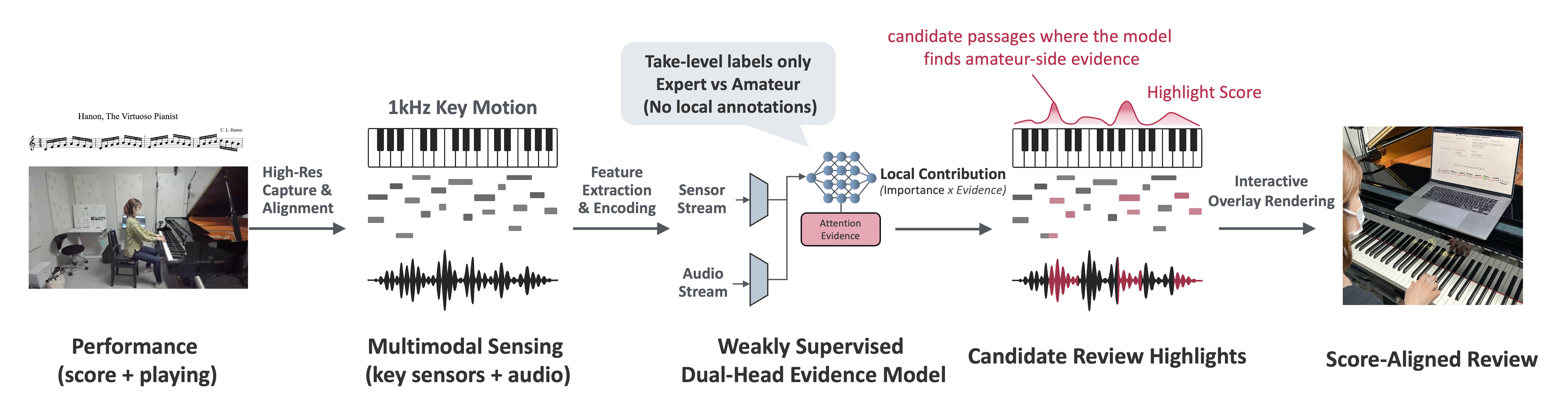}
  \caption{\textbf{Profy: time- and score-aligned highlights for focused review.}
From a single performance, Profy overlays highlights on the timeline and, when alignment is available, on the score to indicate passages for focused review.
The model is trained only with take-level expert/amateur labels, without localized annotations.
It analyzes synchronized 1\,kHz key-motion and audio streams and converts time-localized evidence toward the amateur class into review highlights for scrubbing and looping.}
  \Description{Left-to-right pipeline: a score excerpt and a pianist photo; a panel labeled "Multimodal Sensing" with a keyboard schematic, piano-roll, and waveform; a "Weakly Supervised Dual-Head Temporal Model" note stating "Training labels: expert/amateur only (no localized labels)"; an "Output" panel where notes and waveform segments are shaded red to indicate stronger review highlights; and a final photo showing the score with red highlights. The figure communicates that coarse performance-level labels yield localized, time-aligned visualizations of passages associated with evidence toward the amateur class, with score overlays when alignment is available.}
  \label{fig:teaser}
\end{teaserfigure}

\maketitle

\section{Introduction}

Learning piano performance extends far beyond minimizing incorrect notes. Mastery depends on context-dependent nuances such as micro-timing, articulation, and dynamic balance in tone production~\cite{Lerch2020MPA,Widmer2016ConEspressione}. Although teachers provide contextual guidance, most practice occurs in isolation~\cite{Wilson2023VRPianoReview,McPherson2017SRLMusic}. Consequently, the hardest part of practice is often not execution, but awareness~\cite{HattieTimperley2007,McPherson2017SRLMusic}: learners are often unaware of where, when, and how their playing deviates from pedagogically meaningful expectations. For example, a teacher may focus on a brief thumb-under transition, uneven releases, or a short span of rushing---moments that are easy to miss when practicing alone at fast tempo with both hands. Identifying such moments, rather than simply correcting wrong notes, often becomes the bottleneck for further skill acquisition.

Conventional practice apps and assessment tools often summarize a take with global scores or correctness-oriented metrics~\cite{Eremenko2020AssessmentTech,Kim2022APPA,Wilson2023VRPianoReview}. Such feedback can tell learners that something is off, but rarely where to focus next for practice actions such as scrubbing and looping~\cite{HattieTimperley2007,Kim2022APPA}. Here, scrubbing rapidly moves playback to rehear a short moment, while looping repeats a short span during focused repetition; together with slow practice, these actions help learners inspect and retry a review passage with low search cost~\cite{Zhukov2009EffectivePractising}. While effective for rough correctness of basic motor issues, these metrics capture only a narrow slice of performance quality. A survey of automatic piano performance assessment highlights this gap: most systems output global proficiency scores rather than localized cues for iterative motor-control adjustment~\cite{Kim2022APPA}. Furthermore, high-performing deep learning models often remain opaque, and musicians benefit more from domain-centered explanations than raw model scores~\cite{BryanKinns2024}. However, obtaining dense, time-aligned labels for nuanced technique (e.g., micro-timing or release quality) is expensive and rarely available in everyday practice~\cite{mcfee2018adaptive,Ilse2018}, making fully supervised localization impractical at scale. Purely unsupervised discrepancy detection can surface unusual moments~\cite{Ruff2021AnomalyReview}, but it does not explicitly target the moments most relevant to expert-labeled versus amateur-labeled quality differences.

We address these limitations by combining high-resolution sensing with weakly supervised temporal modeling to visualize embodied nuance. High-resolution key-motion sensing captures per-key motion at 1\,kHz, reflecting control processes---movement, timing, and release---rather than only the acoustic outcome or discrete events captured by standard MIDI sensors~\cite{Oku2022Sensors}. Profy treats each performance as a synchronized sequence of key-motion and audio frames and trains a clip-level classifier using only a binary take-level label: expert-labeled or amateur-labeled~\cite{mcfee2018adaptive,Ilse2018,Pinheiro2015}. The temporal head decomposes this prediction into signed frame contributions by combining each frame's importance with evidence toward the expert- or amateur-labeled side, avoiding granular time-localized annotations that are rarely available in everyday practice. At inference time, contributions toward the amateur-labeled class become time-aligned review highlights. These highlights support scrubbing, one-click looping, and score overlays when alignment is available (Figure~\ref{fig:teaser}), giving learners a time-localized evidence signal for review and replay.

We study short technique exercises as a controlled setting for validating local review cues, rather than longer repertoire with phrase-level or stylistic goals. To enable this weakly supervised approach, we curated a multimodal corpus of short technique studies (scales and arpeggios), comprising synchronized key-motion and audio from 73 pianists and 1{,}083 valid takes for modeling and evaluation. Validating localized outputs raises a distinct challenge: expert pedagogical intuition is often tacit and unwritten~\cite{Tominaga2022TeachingExaggerations}. We therefore designed an annotation workflow to elicit this knowledge as time-localized spans and compare model outputs against human consensus. On a 20-clip subset annotated by 21 expert pianists, the displayed highlight score shows measurable agreement with expert consensus (Pearson $r{=}0.61$, ROC-AUC $0.75$). These results show that behavioral data paired with take-level binary quality labels can yield review cues without dense local annotation. Profy therefore directs attention to review passages for inspection and replay.

The contributions of this paper are as follows.
\begin{enumerate}[topsep=3pt, itemsep=2pt, parsep=0pt]
    \item We introduce Profy, a piano practice system that renders time-aligned review passages for scrubbing, looping, and score-linked review.
    \item We present a weakly supervised multimodal model that decomposes take-level expert/amateur predictions into signed, time-localized evidence scores.
    \item Using 1{,}083 valid synchronized takes from 73 pianists, we evaluate Profy with expert-annotated spans and show alignment with expert judgments.
\end{enumerate}


\section{Related Work}
We situate Profy at the intersection of four threads: (\S\ref{sec-rw-interactive}) interactive piano learning and embodied feedback; (\S\ref{sec-rw-assessment}) automatic music performance assessment and localized feedback; (\S\ref{sec-rw-xai}) weakly supervised, interpretable, and human-centered feedback; and (\S\ref{sec-rw-data}) datasets and sensing modalities for piano performance analysis.
Prior work has demonstrated the value of localized practice feedback.
Profy builds on this direction by learning inspectable timeline review cues from take-level labels, using high-resolution key motion and audio, and exposing those cues for scrubbing, looping, and optional score-linked review.


\subsection{Interactive Piano Learning and Embodied Feedback}
\label{sec-rw-interactive}

Classic systems such as Piano Tutor explored real-time feedback in lesson contexts~\cite{Dannenberg1992}.
Projection and AR interfaces later brought cues onto or around the physical keyboard, including P.I.A.N.O.~\cite{Rogers2014}, OnCall Piano Sensei~\cite{Chiang2015}, and piARno~\cite{Rigby2020}.
Surveys of augmented piano prototypes and AR/VR piano tutors show that many systems support note following, synchronization, or visual guidance, but richer feedback about embodied technique, articulation, timing nuance, and practice planning remains difficult to provide during solo practice~\cite{Deja2022SurveyAugmentedPiano,Wilson2023VRPianoReview}.

Piano-specific systems have moved beyond key-level correctness toward embodied technique visualization.
PianoHandSync visualizes discrepancies between hand postures in two piano performance videos~\cite{Liu2023PianoHandSync}, while PianoSyncAR superimposes time-varying teacher hand poses over the learner's hands in augmented reality~\cite{Liu2023PianoSyncAR}.
PiaMuscle further explores estimating and visualizing miniature hand-muscle activity during piano training~\cite{Liu2025PiaMuscle}.
These systems demonstrate the value of making otherwise tacit technique visible.
However, they typically rely on explicit reference demonstrations, hand-pose comparison, or physiological estimation.
Profy instead localizes review passages from a complete performance using coarse take-level binary quality labels, without requiring a teacher-reference motion for each passage.

Haptic and physiological systems provide complementary forms of practice support.
Mobile Music Touch uses passive tactile stimulation to support piano melody learning~\cite{Kohlsdorf2010MobileMusicTouch}, and Passive Haptic Rehearsal extends this line toward augmented piano learning in everyday settings~\cite{Gemicioglu2024PassiveHaptic}.
Cross-instrument work such as MusicJacket shows how vibrotactile feedback can support instrumental technique learning~\cite{Johnson2010}, while Learn Piano with BACh adapts practice difficulty based on brain state~\cite{Yuksel2016}.
Reflective systems such as EyePiano use gaze to help learners identify difficult passages in the score~\cite{Karolus2023}.
Creative-access and collaboration systems such as Piano Genie and MirrorFugue lower barriers to musical participation and remote co-presence rather than evaluating nuanced technique~\cite{Donahue2019,xiao2011}.

Profy complements these systems by targeting the action layer of practice.
In solo practice, scrubbing, looping, repetition, and slow playback are related but distinct actions: scrubbing helps learners relocate and rehear a moment, looping supports repeated inspection or retry of a short span, and slow playback exposes fine timing and coordination details.
Profy produces localized review cues that can be directly turned into these replay actions while avoiding dense local annotation and per-passage reference demonstrations.


\subsection{Automatic Music Performance Assessment and Localized Feedback}
\label{sec-rw-assessment}

Score--performance alignment and score following are fundamental for practice feedback because they allow performance events to be localized on a timeline or score~\cite{nakamura2016}.
Score-informed assessment has shown that incorporating score information can improve music performance assessment, for example by jointly modeling performance and score-derived structure~\cite{Huang2020ScoreInformedNetworks}.
Benchmarks also frame recorded piano performances as objects of performance understanding rather than transcription alone~\cite{Zhang2024PianoJudges}.
Within music education, surveys of automatic piano performance assessment emphasize that useful feedback should support learner goals, progress tracking, and personalized guidance, not only produce an overall score~\cite{Kim2022APPA}.
Evaluation-oriented datasets such as an expert/novice piano-performance corpus further show the value of combining amateur performances with expert and peer ratings and score alignments for computer-aided feedback~\cite{Jiang2023}.
Many research prototypes and commercial practice tools emphasize score following, correctness feedback, or aggregate summaries.
These are valuable for entry-level practice, but they often leave learners to infer which short passages deserve focused review and replay.

Closer to our goal, localized performance feedback has begun to emerge.
A score-independent conspicuous-mistake detector localizes conspicuous regions in piano performances from MIDI piano-roll data using region-level annotations and a TCN-based detector~\cite{Morsi2023SoundsOutOfPlace}.
A critical-segment analysis approach identifies performance segments that affect piano evaluation using score-derived reference guidelines, expert annotations, SHAP-based feature importance, change-point analysis, and LLM-generated explanations~\cite{Jeon2025CriticalSegments}.
LLaQo further moves toward coaching-oriented assessment by generating query-based formative feedback from audio~\cite{Zhang2025LLaQo}.
These works show that localization and coaching-oriented feedback are an emerging direction rather than an absent one.

Profy therefore differs in its supervision and interaction target.
It learns local review cues from take-level labels only, uses high-resolution key-motion and audio rather than MIDI piano roll alone, and presents time-aligned highlights for scrubbing and looping rather than generating definitive textual assessments.
Unlike score-informed or reference-comparison methods, the timeline highlights do not require a single reference performance or note-wise discrepancy computation, although score alignment can be used when available to render overlays on the staff.

Other work has explored richer representations of expressive performance.
Con Espressione frames expressive performance modeling as a route toward computationally understanding musical nuance~\cite{Widmer2016ConEspressione}.
Real-time abstract visuals of timing and loudness can improve learners' reproduction of expressive details~\cite{Sadakata2008}, and comparative performance visualizations can support analysis across recordings~\cite{Sapp2007Timescapes}.
Automatic evaluation of articulation-specific piano performance, such as legato and staccato quality, further shows promise for piano-specific assessment~\cite{Phanichraksaphong2021}.
However, these systems typically focus on explicit dimensions, fixed references, or global/visual summaries, whereas Profy prioritizes review spans from coarse labels.


\subsection{Weakly Supervised, Interpretable, and Human-Centered Feedback}
\label{sec-rw-xai}

Profy's formulation is related to weakly supervised temporal localization and multiple-instance learning.
In weakly labeled sound event detection, models are trained from clip-level labels while still producing temporally varying predictions that must be pooled during training~\cite{mcfee2018adaptive}.
Attention-based multiple-instance learning similarly treats a labeled bag as a set of instances and learns instance contributions to the bag-level prediction~\cite{Ilse2018}.
Weak supervision has also been used to recover local predictions from image-level labels~\cite{Pinheiro2015}.
Profy applies this design pattern to piano practice: it learns from take-level binary labels (expert-labeled vs.\ amateur-labeled) while producing a time-indexed evidence curve for review.

Because Profy exposes model-derived highlights to learners, the form of explanation matters.
Prior work warns that attention weights alone should not automatically be treated as explanations~\cite{JainWallace2019}, while later work argues that whether attention is explanatory depends on the definition and evaluation of explanation~\cite{WiegreffePinter2019}.
Profy therefore does not present attention weights alone as explanations.
Instead, it visualizes signed additive frame contributions: displayed highlights indicate directional evidence toward the amateur-labeled side of the classifier.
We interpret these outputs as contextual review cues rather than causal diagnoses of performance quality.

In education, AI feedback is more likely to be adopted when it is understandable, supports reflection, and complements teachers rather than replacing them.
Intelligent tutoring and classroom orchestration work emphasizes transparency, teacher--AI complementarity, and actionable insights rather than opaque scores~\cite{Porayska2016,Holstein2019}.
Explainable AI for music similarly stresses musician-centered explanations grounded in musical concepts~\cite{BryanKinns2024}.
Open Learner Models externalize learner state so that learners can inspect, reflect on, and plan from learning data~\cite{BullKay2007OLM,BullKay2016OLM}.
Human-AI interaction guidelines likewise emphasize user control, appropriate trust, and intelligible system behavior~\cite{Amershi2019GuidelinesHAI}.
Profy follows this orientation by acting as an inspectable learner-facing feedback representation rather than an autonomous evaluator that delivers definitive judgments~\cite{Kawamura2021,Kawamura2022DDSupport}.

The broader challenge of turning tacit expertise into situated guidance also appears in medical, vocational, and craft learning.
AR tutoring for machine tasks can adapt instruction to learner task state~\cite{Huang2021AdapTutAR}; HMD-based VR/AR systems support spatial and procedural learning in medical education~\cite{Xu2021HMDMedicalReview}; and mixed-reality or AI systems for origami, wheel throwing, and woodworking explore how to make tacit craft knowledge inspectable~\cite{Chen2025OrigamiSensei,Ji2025ReshapingCraftLearning,Hao2025TacitTeaching}.
These domains motivate the broader relevance of Profy's design pattern, while Profy's technical and interaction contribution is grounded in piano practice: weakly supervised localization of reviewable passages from complete performances.

\subsection{Datasets and Sensing Modalities for Piano Performance Analysis}
\label{sec-rw-data}

Widely used piano corpora such as MAESTRO, MAPS, and ASAP remain foundational for audio/MIDI transcription, alignment, and score-linked analysis~\cite{Hawthorne2019MAESTRO,Emiya2010MAPS,Foscarin2020ASAP}.
Additional datasets extend the design space toward perceptual evaluation and computer-aided feedback.
PercePiano contributes expert-annotated multilevel perceptual evaluation labels for piano performance assessment~\cite{Park2024PercePiano}.
An expert/novice evaluation dataset includes amateur performances, expert and novice ratings, digital scores, and audio-to-score alignments~\cite{Jiang2023}.
These resources are important comparators because they foreground assessment criteria and perceptual evaluation rather than transcription alone.

Multimodal piano datasets also add modalities that are closer to embodied practice.
PianoVAM includes synchronized video, audio, MIDI, hand landmarks, fingering labels, and metadata from amateur practice sessions~\cite{Kim2025PianoVAM}.
PianoMotion10M provides a large-scale benchmark for piano hand-motion generation from bird's-eye videos with annotated hand poses~\cite{Gan2024PianoMotion10M}.
These datasets show increasing interest in multimodal and embodied piano-performance analysis.

High-rate key-motion sensing provides a different view of piano performance: continuous key travel and release dynamics rather than only discrete MIDI events or external hand pose.
A non-contact sensing system with approximately millisecond temporal resolution has been used to identify key-motion features associated with virtuosity~\cite{Oku2022Sensors}; synchronized EEG and key-sensor recordings have linked neural measurements to fine motor timing during piano performance~\cite{Yasuhara2024iScience}; and additional work connects high-resolution key-motion features to perceptual outcomes such as timbre~\cite{Kuromiya2025PNAS}.

Relative to these resources, our corpus is distinctive not simply because it is multimodal, but because it pairs continuous 1\,kHz per-key motion trajectories and audio with take-level binary quality labels, enabling validation of weakly supervised review cues against localized expert spans.
This combination supports the specific problem addressed by Profy: learning and evaluating time-localized review cues without training on dense local annotations.

\section{Dataset and Task Suite}
\label{sec:dataset}

We curated a specialized multimodal corpus to train and evaluate Profy under controlled conditions. Rather than relying on unconstrained recordings, our goal was to capture high-precision measurements of key mechanics alongside audio across a broad cohort, reducing the likelihood that detected expert-labeled versus amateur-labeled differences would be dominated by recording variance. We collected synchronized 1\,kHz per-key motion and 44.1\,kHz audio from 80 pianists recorded under a consistent protocol; after curation, the modeling and evaluation set contains 1{,}083 valid takes from 73 performers. Table~\ref{tab:profy_glance} summarizes the corpus.

\subsection{Capture Protocol and Task Suite}
\label{subsec:protocol}
We recruited eighty pianists spanning amateurs to experienced performers and teachers. These background descriptors characterize recruitment, while the modeling target is assigned per take from listener ratings. Performances were captured on a digital piano instrumented with the HackKey system, a non-contact optical sensing setup using photo reflectors (LBR-127HLD) to measure the vertical position of all 88 keys at 1\,kHz (1\,ms resolution)~\cite{Oku2022Sensors}. Following prior HackKey descriptions, the sensing pipeline is calibrated per key using neutral and bottom positions, enabling conversion to physical key displacement in millimeters. Audio was recorded at 44.1\,kHz as stereo PCM using a Zoom H6 recorder with an XYH-6 capsule, with a fixed placement 0.5\,m above the keyboard center and 0.4\,m in front of the instrument in a quiet room. We chose microphone capture rather than direct line-out to reflect practice-like acoustic conditions. We used a fixed microphone placement and recording chain throughout the corpus. Both streams are timestamped at capture, but small start offsets can remain due to device latencies. Before modeling, we estimate a per-take offset from onset detection and energy cross-correlation, align the recorded streams on a common timeline, and interpolate them to the fixed-length model time base used downstream.

\begin{figure*}[t]
  \centering
  \includegraphics[width=0.80\textwidth]{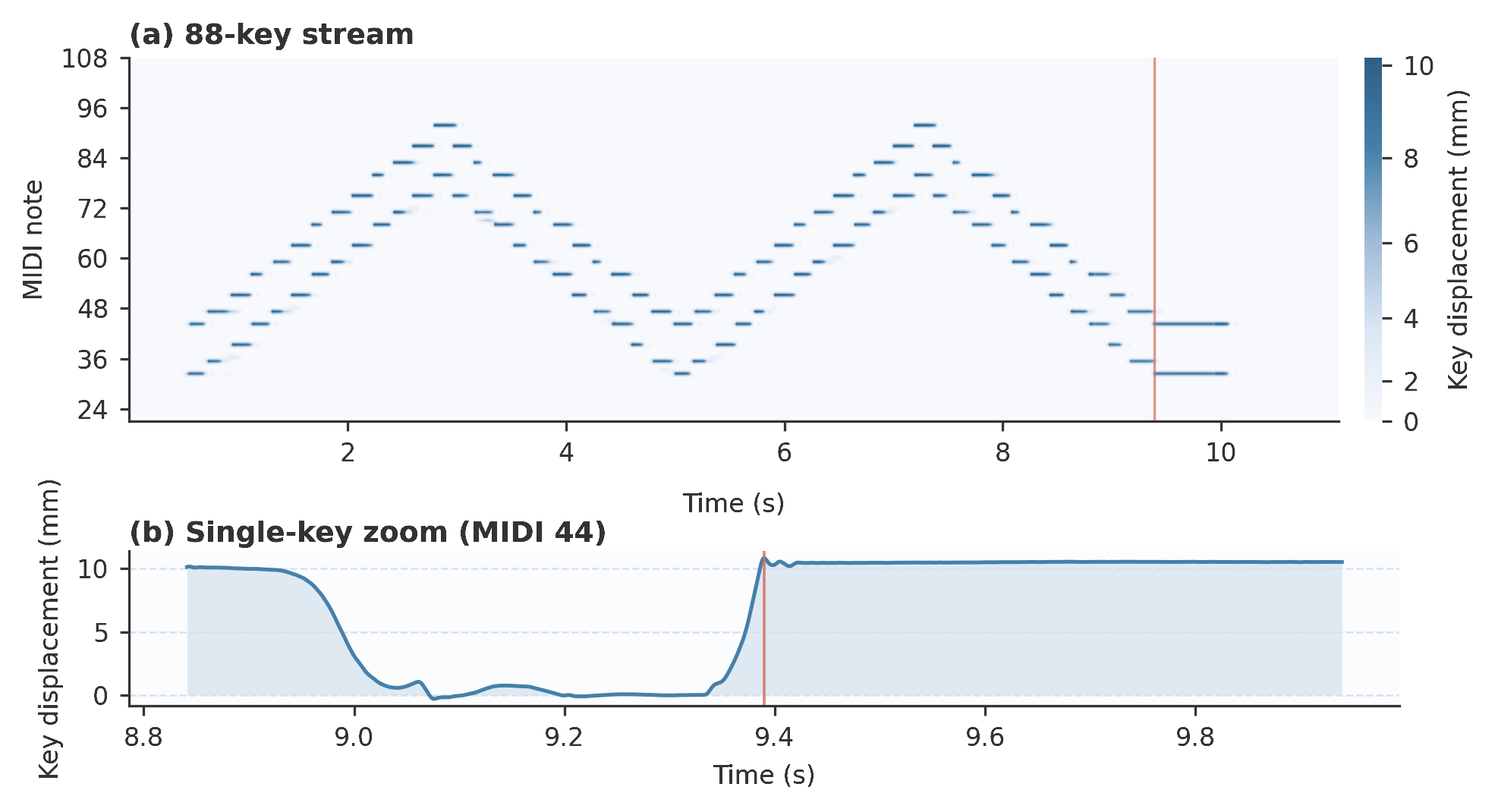}
  \caption{\textbf{Example raw key-motion sensor data.} Each take provides a 1\,kHz time column and 88 calibrated key-displacement channels (mm). We visualize the full 88-key stream as a time-by-key heatmap, along with a zoomed view of a single channel.}
  \Description{A two-panel figure. The top panel is a heatmap of calibrated key displacement (mm) over time for all 88 keys. The bottom panel is a line plot of a single key channel over a short window, with a vertical line marking a salient moment.}
  \label{fig:sensor_sample}
\end{figure*}

The raw key-motion data provide a 1\,kHz time base and 88 per-key vertical displacement channels in millimeters, derived from per-key neutral/bottom calibration. Figure~\ref{fig:sensor_sample} shows an example from a single take.

The task suite consists of fifteen short studies designed to stress specific motor control challenges where amateur instability typically manifests.
We selected nine scales to probe evenness in timing and loudness during high-velocity finger crossings (thumb-under/finger-over), including keys with varying black-key topography.
Six arpeggios were included to demand larger hand relocations and span management, isolating distinct coordination issues from scalar runs.
The specific keys cover B, C, D$\flat$, E$\flat$, and G$\flat$ major together with B$\flat$, C$\sharp$, E$\flat$, and F$\sharp$ minor (scales), and A$\flat$, D, and F major together with B$\flat$, F, and G$\sharp$ minor (arpeggios).

The recording protocol was uniform. After brief instructions and an optional warmup, each participant could record up to ten takes per task. Across participant--task pairs, the median number of recorded takes is one and the mean is 2.04. Tempo was self-selected unless otherwise indicated. Participants were asked to aim for steady timing and even dynamics rather than concert speed.

\begin{table}[t]
  \centering
  \caption{\textbf{Corpus at a glance. All values are computed from the dataset; valid denotes retained recorded takes with complete sensor/audio used for modeling. Label statistics additionally report the broader rated subset from which the valid modeling set is derived.}}
  \Description{Two-column summary table of the dataset. It reports 80 participants, 73 valid participants, 2,433 total recorded takes, 1,175 retained takes, and 1,083 valid takes across 15 tasks. It also lists average takes per piece, take duration statistics, the two recording modalities, and performance-level Expert versus Amateur take labels derived from 6,517 ratings provided by 53 raters, covering 1,084 rated takes in total and 1,083 valid takes used for modeling.}
  \label{tab:profy_glance}
  \begin{tabular}{@{}>{\raggedright\arraybackslash}p{0.30\columnwidth}>{\raggedright\arraybackslash}p{0.64\columnwidth}@{}}
    \toprule
    Item & Value \\
    \midrule
    Participants (unique / valid) & 80 / 73 \\
    Recorded takes (total / retained / valid) & 2{,}433 / 1{,}175 / 1{,}083 \\
    Piece tasks (unique) & 15 (9 scales, 6 arpeggios) \\
    Takes per task per participant (mean / median / max) & 2.04 / 1 / 10 \\
    Take length (valid; median [min, max]) & 10.72\,s [7.86, 16.99] \\
    Modalities & 1\,kHz per-key vertical displacement (mm; non-contact optical, nominal 0.01\,mm resolution); 44.1\,kHz stereo audio (resampled to 22.05\,kHz for features) \\
    Labels & Expert-labeled / amateur-labeled take labels derived from median-thresholded aggregated listener ratings; 6{,}517 ratings from 53 raters (1{,}084 rated takes; 1{,}083 valid for modeling) \\
    \bottomrule
  \end{tabular}
\end{table}

\subsection{Labels, Curation, Contents, Uses, and Limitations}
\label{subsec:labels-content}

Each take is assigned a binary performance-level label (expert-labeled or amateur-labeled) from aggregated listener ratings. Each rating provides two scalar assessments: technique and musicality. For take $i$, we compute take-level means over the available raters $R_i$,
\[
\begin{aligned}
\mathrm{technique}_i
&=\operatorname{mean}_{r\in R_i}\mathrm{technique}_{ir},\\
\mathrm{musicality}_i
&=\operatorname{mean}_{r\in R_i}\mathrm{musicality}_{ir},
\end{aligned}
\]
and combine them as
\[
\mathrm{quality}_i = 0.5 \, (\mathrm{technique}_i + \mathrm{musicality}_i).
\]
We then derive the binary performance label by thresholding the combined mean score at the corpus median over rated takes:
\[
y_i=\mathbb{I}\!\left[\mathrm{quality}_i \ge \operatorname{median}_{j}\mathrm{quality}_j\right],
\]
with $y_i{=}1$ reported as expert-labeled and $y_i{=}0$ as amateur-labeled.

Across the corpus, 53 independent raters provided 6{,}517 ratings. In total, 1{,}084 takes received ratings; intersecting this rated subset with retained takes and complete sensor/audio capture yields 1{,}083 valid takes from 73 performers for modeling. The resulting valid set contains 597 expert-labeled and 486 amateur-labeled takes. Because the label rule uses $\ge$ at the median and the valid set is the intersection with retained complete captures, the valid set is not exactly balanced. Continuous ratings are not used directly as regression targets; only the binarized labels derived from the aggregated take-level means are used for training. No localized annotations are used for model training.

Curation was conservative. A recorded take was retained only if all metadata records were complete, capture was intact, the protocol was followed, and the content was valid for the declared task. Of 2{,}433 recorded takes, 1{,}175 were retained. During our modeling pipeline, we compute indicators of acoustic reliability, including non-silence ratio, spectral flatness, and loudness, and use them to drive recording-quality-aware fusion and ablations.

Each recorded take includes take-level descriptors and two time-stamped streams: 1\,kHz key-motion and 44.1\,kHz stereo audio.
We also extract per-note onset/offset times and kinematic summaries from the key-motion stream for analysis. We define a key-press event when the key displacement first exceeds 5\,mm from the neutral position (as in prior HackKey-based setups~\cite{Yasuhara2024iScience}), and a key-release event when it returns below 3\,mm. Key-motion values are per-key vertical displacements in millimeters, obtained by calibration between neutral and bottom positions. All time bases are in seconds, and frame rates are explicitly recorded with each take.

We use this corpus for three evaluation tasks in \S\ref{sec:technical-eval}: expert-versus-amateur classification, weakly supervised localization, and robustness under input corruption. When a score representation and reliable score following are available~\cite{nakamura2016}, evidence learned from weak supervision can be aggregated to notes or beats and rendered on the staff. Otherwise, we use timeline overlays aligned by absolute time.

\paragraph{Ethics, consent, and data handling.}
All study procedures were reviewed and approved by the Sony Ethics Committee (approval no.\ 24-25-0001).
All participants provided informed consent prior to data collection.
We stored all recordings and sensor traces in de-identified form with access restricted to the research team, and we do not report any personally identifying information.

\begin{figure*}[t]
  \centering
  \includegraphics[width=0.9\linewidth]{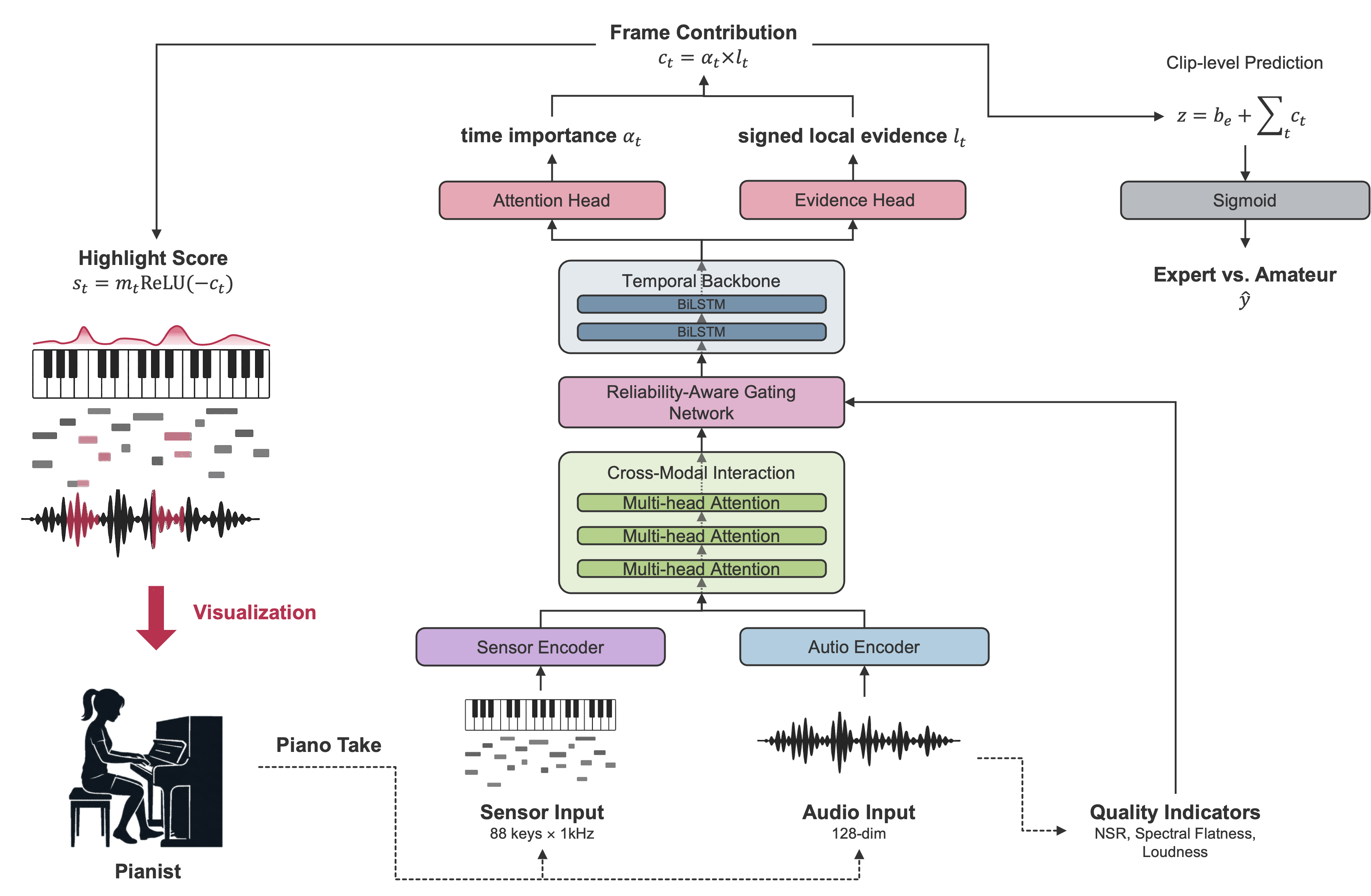}
\caption{\textbf{Profy architecture.} Sensor and audio streams are encoded separately, exchanged through 4-head bidirectional cross-modal attention, and fused by a reliability-aware gating network using acoustic quality indicators. A 2-layer BiLSTM feeds an attention head and an evidence head. The attention head estimates frame importance $\alpha_t$, while the evidence head estimates signed local evidence $\ell_t$. The clip-level prediction uses frame contributions $c_t=\alpha_t\ell_t$ as $z=b_e+\sum_t c_t$; the UI highlights contributions toward the amateur-labeled class, $s_t=m_t\mathrm{ReLU}(-c_t)$, as review spans.}
  \Description{Architecture diagram of the Profy model showing sensor and audio encoding, cross-modal exchange, reliability-aware fusion, a temporal backbone, and two output heads. The attention head produces frame importance weights, the evidence head produces signed local evidence, and their product defines frame contributions used for both clip prediction and highlight visualization toward the amateur-labeled class.}
  \label{fig:architecture_overview}
\end{figure*}


\section{Profy: Methodology and Architecture}
\label{sec:profy_method}

\subsection{Approach Overview}
\label{subsec:overview}

Profy's goal is to help learners decide where to focus practice by making expertise-dependent expression differences inspectable.
Given a single piano take, Profy analyzes synchronized key-motion and audio streams and identifies moments whose learned evidence points toward the amateur-labeled rather than expert-labeled side of performance quality.
The interface then visualizes these moments as time-aligned highlights on the timeline and, when alignment is available, on the score, so learners can listen, scrub, and loop the passages most worth reviewing.
The core constraint is that we only have take-level supervision (expert-labeled vs.\ amateur-labeled), not per-note or per-span labels.
A second constraint is interaction: highlights must be stable enough to support repeated playback, looping, and (when available) score overlays.
Finally, audio quality varies across takes (e.g., silence and room noise), so the system is designed to remain informative under varying recording conditions.

Our pipeline (Fig.~\ref{fig:architecture_overview}) has three stages.
First, we align sensor and audio streams and extract synchronized features.
Second, we encode each modality and fuse them with reliability-aware gating.
Third, we predict a clip-level expert-labeled versus amateur-labeled probability while simultaneously producing a signed, time-localized contribution curve that we use for visualization.
Accordingly, the displayed highlights are derived not from attention alone but from each frame's directional contribution to the final expert-labeled versus amateur-labeled prediction.
Intuitively, the model separates two questions that are often conflated in saliency visualizations: which moments matter for the clip-level decision, and which side of the decision each moment supports.
The attention head answers the first question by assigning a time-importance weight to each model frame, while the evidence head answers the second by estimating whether that frame supports the expert-labeled or amateur-labeled side.
The interface visualizes frames that are both decision-relevant and supportive of the amateur-labeled class as review cues.
The resulting highlight suggests where a learner may listen, loop, and inspect next.

\subsection{Inputs and Interface Outputs}
\label{subsec:io}

Each take provides two synchronized streams:
(1) a sensor feature sequence $\mathbf{S}\in\mathbb{R}^{T\times D_s}$ derived from 1\,kHz per-key motion, and
(2) an audio feature sequence $\mathbf{A}\in\mathbb{R}^{T\times 128}$ computed from the microphone recording.
The audio sequence is obtained by concatenating frame-level descriptors and projecting them to the 128-dimensional representation consumed by the model.
We also compute an audio non-silence mask $\mathbf{m}\in\{0,1\}^{T}$ and a compact quality vector $\mathbf{q}\in\mathbb{R}^{3}$ (NSR, spectral flatness, loudness) to inform reliability-aware fusion.
Training uses only the take label $y\in\{0,1\}$, where $0$ denotes the amateur-labeled class and $1$ denotes the expert-labeled class.

At inference time, Profy produces a small set of signals used by the interface:
(1) an expert-labeled probability $\hat{y}$,
(2) a time-series highlight score $s_{1:T}$ used to render overlays,
(3) modality weights that indicate how much each modality contributed to the fused representation (for reliability cues), and
(4) quality indicators $(\mathbf{m},\mathbf{q})$ so the UI can suppress overlays when inputs are untrustworthy.

\subsection{Preprocessing and Leakage Controls}
\label{subsec:preproc}

We map each take to a fixed-length time base $T{=}1000$ to keep the visualization and interaction consistent across variable durations.
For each take, we first estimate a sensor--audio offset (onset detection + energy cross-correlation), then align the recorded streams on a common timeline, and finally interpolate both streams to the shared model time base of length $T$. In this pipeline, the audio frame grid serves as the intermediate alignment reference, whereas $T{=}1000$ is the final time base used by all downstream encoders.
Throughout the remainder of the method, a ``frame'' refers to an index on this shared resampled model time base rather than to a native 1\,ms sensor sample.
For sensors, $\mathbf{S}$ contains key-motion summaries (e.g., position/velocity/acceleration and event indicators) aggregated across 88 keys.
For audio, raw log-Mel, MFCC, chroma, tonnetz, and spectral descriptors are concatenated per audio frame and projected to the 128-dimensional descriptor sequence $\mathbf{A}$.
We compute (i) a non-silence mask $\mathbf{m}$ from short-term energy and (ii) $\mathbf{q}=[\mathrm{NSR},\mathrm{spectral\ flatness},\mathrm{loudness}]$ to represent acoustic reliability.

To avoid train--test leakage, all feature standardization is performed per fold using statistics from the training split only (restricted to non-silent frames for audio), and then applied to validation/test data in that fold.

\subsection{Model: Reliability-aware Fusion + Contribution-based Highlighting}
\label{subsec:arch}

Profy uses the following highlighting rule: a model-frame state is highlighted when its contextual representation contributes evidence toward the amateur-labeled class.
To make this explicit, we structure the temporal head so that the clip-level prediction can be decomposed into signed per-frame contributions.

\subsubsection{Multimodal encoding and cross-modal exchange}

Two lightweight encoders map $\mathbf{S}$ and $\mathbf{A}$ into sequences in a shared hidden space ($d{=}256$):
$\mathbf{H}^{S}\in\mathbb{R}^{T\times d}$ and $\mathbf{H}^{A}\in\mathbb{R}^{T\times d}$.
We use bidirectional cross-attention so that each modality can query the other (motor control $\leftrightarrow$ acoustic outcome)~\cite{Vaswani2017,Lu2019ViLBERT}.
In implementation, the cross-attention module uses four heads, and the branch-to-branch exchanged sequences are returned to the common length with a parametric resampler that combines adaptive average pooling with depthwise and pointwise temporal convolutions.
This yields attended sequences that expose each stream to complementary information while retaining a time-aligned representation.

\subsubsection{Reliability-aware gating for fusion}
\label{subsec:gating}

Audio quality varies substantially across real recordings.
To improve robustness of multimodal fusion, we mix multiple candidate sequences using a small gating network conditioned on the acoustic quality vector $\mathbf{q}$ (and the estimated alignment offset when available).
The gate does not receive an explicit sensor-dropout indicator; changes in sensor share under sensor corruption therefore reflect learned fusion behavior after shared preprocessing rather than direct sensor-failure detection.
Concretely, we form four candidates at this common length: the sensor encoding, audio-to-sensor attended features, the audio encoding, and sensor-to-audio attended features.
The gate outputs mixture weights $w_{1:4}$ (softmax), and the fused representation is a convex combination of candidates.
In implementation, the softmax temperature is increased as NSR decreases.
We also add an NSR-dependent pre-softmax bias toward sensor-side candidates, calibrated to favor a nominal total sensor share of about 0.60 at very low NSR and about 0.25 at high NSR.
This term acts as a soft prior rather than a hard lower bound; the learned gate can override it, so the observed clean-condition sensor share may be higher than the nominal high-NSR value.
This fusion behavior functions as a reliability-aware mixture-of-experts controller~\cite{Jacobs1991MoE,Shazeer2017MoE}.

\subsubsection{Decision decomposition for time-localized evidence}

Let $\mathbf{h}_t$ be the temporal backbone state at model frame $t$ on the shared resampled time base (we use a 2-layer BiLSTM for temporal context).
From each $\mathbf{h}_t$, we predict:
(i) a signed frame-evidence score $\ell_t$ computed without a classifier bias term (positive values support the expert-labeled class, whereas negative values support the amateur-labeled class), and
(ii) a pooling logit $a_t$ that determines how much this frame participates in the clip decision.
\[
\ell_t=\mathbf{w}_e^\top \mathbf{h}_t,\qquad
a_t=\mathbf{w}_\alpha^\top \mathbf{h}_t+b_\alpha.
\]
We convert pooling logits into normalized weights with a masked softmax over non-silent frames:
\[
\alpha_t=\frac{m_t\exp(a_t)}{\sum_{j=1}^{T}m_j\exp(a_j)},\qquad \sum_{t=1}^{T}\alpha_t=1.
\]
If the non-silence mask is all zero, the implementation falls back to an all-ones mask for the softmax and suppresses overlays through the quality indicators passed to the UI.
The clip-level logit is an additive aggregation:
\[
c_t=\alpha_t \ell_t,\qquad
z=b_e+\sum_{t=1}^{T}c_t,\qquad \hat{y}=\sigma(z).
\]
This structure is deliberate: it makes the clip decision decomposable into per-frame contributions $c_t$, while treating $b_e$ as a global classifier bias that is not visualized as frame-local evidence.
In plain terms, the two factors answer different questions.
The weight $\alpha_t$ says how much the model relied on this moment, while the evidence value $\ell_t$ says whether that moment looks more consistent with the expert-labeled or amateur-labeled side.
A high $\alpha_t$ is therefore not enough to decide that a learner should review the span: an important moment may be important because it supports the expert-labeled side, or because it supports the amateur-labeled side.
The product $c_t=\alpha_t\ell_t$ combines these two pieces of information.
Positive values support the expert-labeled side of the model's decision, whereas negative values pull the decision toward the amateur-labeled side.
Because $\mathbf{h}_t$ comes from a BiLSTM and summarizes surrounding context, we interpret $c_t$ operationally as a time-indexed contribution to the model's internal score before probability conversion, rather than as a causal attribution to only the raw input samples at time $t$.

For visualization, we highlight moments that contribute to the amateur-labeled side of the classifier by taking the negative part of the contribution:
\[
s_t = m_t \,\mathrm{ReLU}(-c_t).
\]
This rule means that the interface shows only moments that are both important to the model and directed toward the amateur-labeled side.
For example, consider two non-masked moments, so $m_t{=}1$ and $s_t=\mathrm{ReLU}(-c_t)$.
Suppose both moments have the same importance, $\alpha_t{=}0.25$.
If the first moment has $\ell_t{=}2.0$, then $c_t=\alpha_t\ell_t=0.25\times2.0=0.50$.
This positive contribution supports the expert-labeled side, so $-c_t=-0.50$ and the displayed highlight is $s_t=\mathrm{ReLU}(-0.50)=0$.
If the second moment has $\ell_t{=}{-}2.0$, then $c_t=0.25\times(-2.0)=-0.50$.
This negative contribution points toward the amateur-labeled side, so $-c_t=0.50$ and the displayed highlight is $s_t=\mathrm{ReLU}(0.50)=0.50$.
Thus, Profy's visualization is not simply an attention map of ``important'' moments.
It is a direction-aware review signal: instead of showing only a global score, Profy presents time-localized spans as candidate practice targets---places to listen to, scrub, and loop when deciding where to focus practice.

\subsection{Training with Take-level Labels}
\label{subsec:weak}

All neural parameters (encoders, cross-attention, fusion gate, temporal backbone, and the two linear heads) are trained end-to-end.
Feature extraction and alignment are deterministic preprocessing and are not learned.
We train using only the take label $y$ with a clip-level classification loss:
\[
\mathcal{L}_{\mathrm{cls}}=\mathrm{BCE}\big(y,\hat{y}\big).
\]
Neither head receives direct frame-level supervision: there are no labels indicating which frames should receive high attention or which frames should be locally marked for review.
Instead, the two heads are learned jointly through the clip-level loss: gradients encourage the attention head to place weight on frames whose evidence helps predict the take label, and encourage the evidence head to assign the appropriate direction to those attended frames.
For amateur-labeled takes, this encourages attended frames to provide negative evidence; for expert-labeled takes, it encourages attended frames to provide positive evidence.
This is the mechanism by which take-level supervision can yield a time-localized review signal.
To prevent degenerate solutions and stabilize multimodal behavior, we add two light regularizers that remain compatible with take-level supervision:
\[
\mathcal{L}
= \mathcal{L}_{\mathrm{cls}}
- \lambda_{\mathrm{ent}}\mathcal{H}(w)
+ \lambda_{\mathrm{loc}}\mathcal{R}(s).
\]
$\mathcal{H}(w)$ denotes the entropy of the fusion weights; the negative sign rewards non-collapsed fusion (higher entropy mixtures early in training) under loss minimization. In our implementation, $\mathcal{R}(s)$ combines an $\ell_1$ sparsity penalty on the raw highlight curve with a mild top-$K$ peakiness reward:
\[
\mathcal{R}(s)=\frac{1}{T}\sum_{t=1}^{T}s_t-\rho\left(\frac{1}{K}\sum_{t\in\mathrm{TopK}(s)}s_t-\frac{1}{T}\sum_{t=1}^{T}s_t\right),
\]
where $K=\lceil0.1T\rceil$ and $\rho$ is a small weighting coefficient.
These terms do not require any localized annotations; they shape the form of the highlight signal while keeping the supervision strictly at the take level.

\subsection{From Inference to UI: Evidence-to-Overlays}
\label{subsec:viz}

At inference time, we compute $\hat{y}$ and the raw highlight curve $s_{1:T}$.
For display, we apply display-only post-processing that is not part of learning: clip-wise normalization and optional short-window smoothing.
The interface uses clip-wise normalization because its primary goal is to prioritize passages within the current take.
Clip-wise normalization is intentionally matched to the central practice decision within a take: which passage to inspect next. It therefore prioritizes local review targets within the current performance, while cross-session reflection can be supported through session-to-session deltas or side-by-side comparison views rather than by directly comparing raw within-clip magnitudes.
When all raw highlight values in a clip are identical, min--max normalization returns an all-zero displayed highlight curve to avoid arbitrary peaks.
We then optionally smooth with a short moving average to reduce flicker, and map frame indices back to wall-clock time using the stored resampling grid.
By default we render highlights on a timeline; when score-following is available, we aggregate $s_t$ to notes/beats via the frame-to-score map and render it on the staff.

Finally, we pass $(\mathbf{m},\mathbf{q})$ and the modality weights to the UI so the interface can communicate input reliability, such as low non-silence ratio or high acoustic noise.

\subsection{Splits and Reproducibility}
\label{subsec:repro}

We evaluate with performer-disjoint cross-validation (no performer overlap between folds) and reserve a performer-disjoint validation split within each training fold for early stopping.
We train with AdamW and early-stop on validation Macro-F1.
All preprocessing and feature standardization are performed within each fold to avoid leakage, and we fix random seeds for reproducibility.


\section{Evaluation: Validating Time-aligned Practice Targets Against Human Expertise}
\label{sec:technical-eval}

Profy is built to support practice review by producing time-aligned highlights that point to passages worth revisiting.
This section validates that the model outputs behave as intended for that use.
We structure the evaluation around three questions.
E1 tests whether the model separates expert-labeled and amateur-labeled takes under performer-disjoint evaluation, serving as a preliminary validation of the learning signal.
E2 is the main evaluation: it tests whether the highlight score aligns with expert listeners' review-span markings collected in a separate annotation study, indicating that weakly supervised localization remains coherent.
E3 evaluates whether the displayed highlights remain stable under input corruption.


\subsection{Evaluation Setup}
\label{subsec:eval-setup}

\paragraph{Main corpus and splits.}
We use the 1{,}083 valid takes from 73 performers described in \S\ref{sec:dataset}.
All clip-level results use 3-fold cross-validation grouped by performer so that no performer appears in multiple folds.
Within each training split, we reserve 15\% for validation using performer-grouped sampling.
We report Macro-F1 and accuracy as mean $\pm$ SD over folds.

\paragraph{Models compared and positioning.}
We compare four models: (1) sensor-only, (2) audio-only, (3) a decision-level product-of-experts baseline, and (4) our proposed multimodal fusion with recording-quality-aware gating.
Sensor-only and audio-only are unimodal variants used as ablations to isolate modality contributions.
They share the same temporal attention and evidence heads as the multimodal model, while retaining the same alignment, masking, and fixed-length interpolation pipeline used to construct the common $T$-length representation; the difference is the input stream and encoder used at the prediction stage. We therefore treat them as single-encoder controls on a shared preprocessing stack rather than as preprocessing-isolated end-to-end pipelines.
The decision-level product-of-experts (PoE) baseline combines the clip-level posteriors of separately trained sensor-only and audio-only predictors as $p(y\mid S,A)\propto p_S(y\mid S)\,p_A(y\mid A)$. We report PoE in E1 as a clip-level reference, but not in E2/E3 because it does not produce a single fused frame-level highlight curve or fusion weights comparable to the other models.
These comparisons isolate the effects of multimodal fusion, time-localized highlighting from take-level labels, and reliability-aware behavior under variable input quality.

\paragraph{Expert highlight annotation study for E2.}
To evaluate highlight validity, we conducted a separate annotation study focused on localizing passages that experts selected for review.
We randomly sampled 20 amateur-labeled performances from the corpus and recruited 21 expert annotators to listen and mark review spans using a web-based tool.
These localized span annotations were collected specifically for this evaluation and are not part of the corpus labels used for training.
The model is trained only with take-level binary quality labels.

\begin{figure*}[t]
  \centering
  \includegraphics[width=0.8\linewidth]{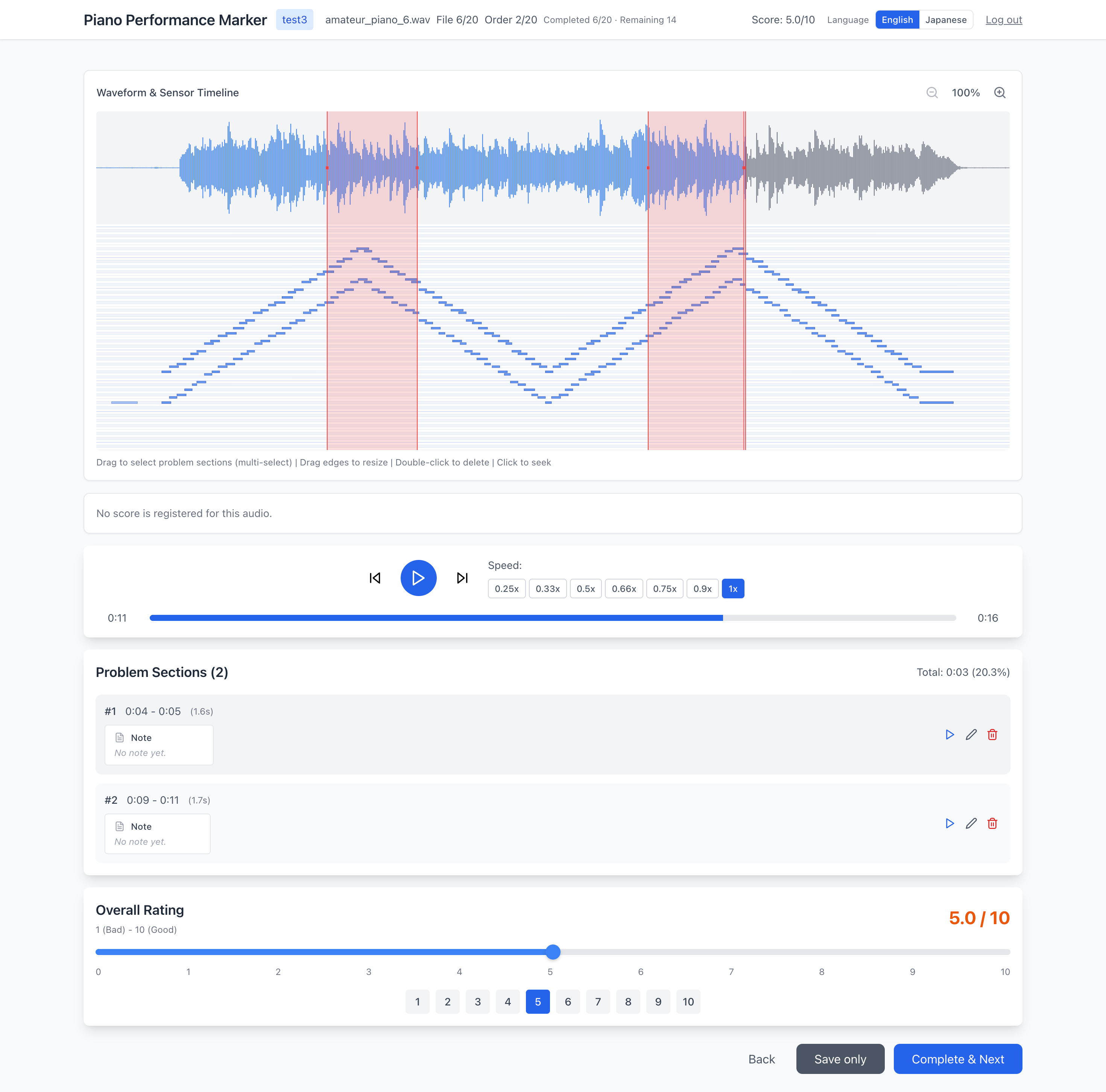}
  \caption{\textbf{Expert annotation tool for localizing review passages.}
  Annotators listen and mark passages for review as editable time spans on a stacked waveform + sensor timeline (piano-roll-style key activity).
  The interface supports scrubbing/zoom, looped playback, and variable playback speed (including slow playback), and allows optional free-text notes per span.
  When a score is available, the tool additionally shows a compact score cue as a location reference.}
  \Description{Screenshot of the web UI used to collect expert annotations, showing playback controls and segment selection.}
  \label{fig:annotation_ui}
\end{figure*}

\paragraph{Annotation interface.}
Because our system output is time-aligned, the annotation interface is designed to capture expert judgments in the same form: time spans.
Annotators work on a stacked waveform and key-motion timeline with zoom and scrubbing, drag to create spans, and adjust boundaries via handles or list edits; spans can also be deleted and replayed in
isolation from the markings list.
To support fine-grained localization of short micro-events, the tool provides (i) a compact score cue shown under the waveform as a location reference for the exercise, (ii) playback speed control
(including slow playback), and (iii) an optional free-text note field attached to each span for brief pedagogical rationales (e.g., thumb-under coordination, persistent inter-hand drift).
We randomized clip order per annotator and included a completion check that surfaces any unanswered clips before submission.
Figure~\ref{fig:annotation_ui} shows the annotation interface used in the main study.

\paragraph{Formative pilot and iterative refinement.}
To refine the annotation protocol, we first conducted a formative pilot with eight professional pianists, asking them to annotate a subset of ten randomly selected takes.
Our initial interface design relied solely on a waveform visualization for navigation.
However, pilot participants reported that identifying specific technical issues on a waveform was cognitively demanding, often leading to fatigue and coarser span selections.
Specifically, they noted that (a) waveform peaks do not map semantically to musical structure, making it hard to locate specific notes, and (b) standard playback speed obscures micro-timing deviations.
Consequently, we refined the tool to include a score-based location reference for structural anchoring and variable playback speed to support inspection of rapid motor details.
Furthermore, participants expressed a desire to explain the nature of the issue rather than just its location; thus, we added free-text rationale fields.
We also introduced procedural safeguards, including randomized clip ordering to mitigate fatigue effects, which were finalized in the protocol used for the main study.
Pilot annotations are excluded from all reported agreement metrics.

\begin{figure*}[t]
  \centering
  \includegraphics[width=0.92\textwidth]{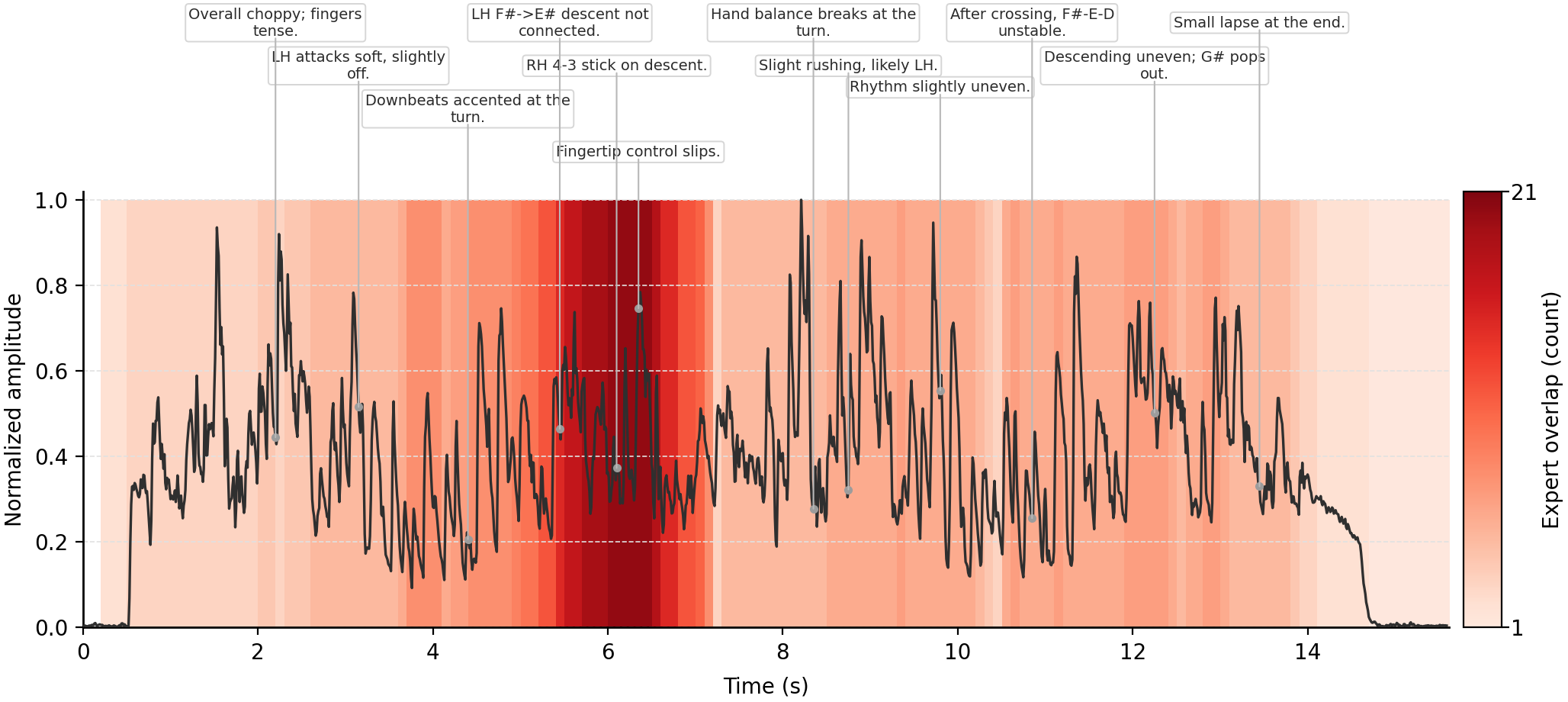}
  \caption{\textbf{Example overlay of expert-marked spans and comments on an amateur-labeled clip.} The background heatmap encodes expert overlap over time (darker = higher agreement), and callouts show the attached free-text notes.
  The black trace provides a normalized signal reference for timing.}
  \Description{Timeline plot with a heatmap of expert overlap and callouts containing short textual comments placed above the timeline.}
  \label{fig:annotation_overlay}
\end{figure*}

\paragraph{From spans to a soft consensus curve.}
For each clip, we convert marked spans to a frame-level vote proportion $v_t\in[0,1]$,
defined as the fraction of eligible annotators whose spans cover frame $t$.
We include only annotators who completed the full randomly sampled 20-clip set.
To reduce isolated marks that do not reflect shared expert judgment, we suppress frames supported by fewer than four votes.
Because span boundaries are approximate at sub-second resolution, we expand each span by an annotation margin of 0.5\,s before voting.
We then apply a mild gamma compression ($\gamma{=}0.7$) to obtain the final soft consensus curve $g_t = v_t^\gamma$.
For thresholded metrics, we binarize the consensus by selecting a per-clip threshold that yields 20\% positive coverage,
producing a binary mask $G_t \in \{0,1\}$.
Figure~\ref{fig:annotation_overlay} shows an example overlay of expert spans and attached notes.

Unless otherwise noted, E2 uses a 0.5\,s span margin, a minimum of four votes, gamma compression with $\gamma{=}0.7$, and a threshold yielding 20\% positive coverage; note capture uses a 30\% span-overlap criterion. These values define a fixed evaluation operating point used for comparability across clips. This operating point is chosen to reflect realistic practice-oriented review. The vote threshold suppresses isolated markings, the span margin absorbs sub-second boundary uncertainty, gamma compression preserves shared peaks, and 20\% positive coverage yields sparse review units that remain easy to inspect through replay. The 30\% span-overlap rule credits partial localization of short expert comments without rewarding incidental overlap. We use this single operating point to keep E2 directly comparable across clips.

\paragraph{Highlight score used for evaluation.}
We evaluate the same clip-normalized highlight score that the UI displays (\S\ref{subsec:viz}). The score is used to rank passages within a take for review.
Specifically, given attention weights $\alpha_t$ and signed frame-evidence scores $\ell_t$ computed without a classifier bias term (\S\ref{subsec:arch}),
we compute per-frame contributions $c_t = \alpha_t \ell_t$, excluding the global classifier bias from visualization, and define the raw highlight score as
\[
s_t = m_t \,\mathrm{ReLU}(-c_t),
\]
which highlights frames that push the clip-level prediction toward the amateur-labeled class.
Unless otherwise noted, all E2 metrics are computed on the per-clip normalized highlight score $\tilde{s}_t$
obtained by min--max normalization of $s_t$ within each clip.
If all $s_t$ values are identical within a clip, we set $\tilde{s}_t=0$ for all frames.
For the multimodal model, $\alpha_t$ and $\ell_t$ are taken from the multimodal head operating on the fused, gate-weighted representation.


\subsection{E1: Clip-level Expertise Classification}
\label{subsec:eval-classification}

We first report clip-level expert-labeled versus amateur-labeled classification under performer-disjoint cross-validation.
This analysis examines whether the corpus contains a learnable expertise signal and whether Profy's model can recover that signal for performers unseen during training.
Because the valid set is mildly imbalanced, we emphasize Macro-F1 alongside accuracy.
Strong clip-level performance also supports the later use of the model's signed evidence curve for time-localized review cues.

Table~\ref{tab:main_results} shows that all learned variants exceed the majority-class baseline.
We include that deterministic baseline for reference.
Sensor-only and audio-only models reach similar performance, indicating that both key-motion and audio streams contain learnable expertise-related information.
The decision-level PoE baseline, which combines separately trained unimodal posteriors, does not outperform the learned multimodal fusion model.
The multimodal model achieves the highest Macro-F1 and accuracy, suggesting that reliability-aware fusion can use the two streams jointly rather than merely combining independent decisions.
These results establish that the take-level labels provide a learnable signal under strict splits; E2 then evaluates whether the resulting local highlights align with expert-marked review spans.

\begin{table}[!t]
  \centering
  \caption{Cross-validated performance on expert-labeled versus amateur-labeled classification using 3-fold performer-disjoint splits. Values are mean $\pm$ SD over folds; the majority-class baseline is deterministic.}
  \Description{Performance table with four learned models and one baseline. The majority-class baseline has Macro-F1 0.355 and accuracy 0.551. Sensor-only reaches Macro-F1 0.756 and accuracy 0.775, audio-only 0.759 and 0.769, decision PoE 0.753 and 0.772, and the multimodal model performs best with Macro-F1 0.781 and accuracy 0.782.}
  \label{tab:main_results}
  \begin{tabular}{@{}lcc@{}}
    \toprule
    Mode & Macro-F1 & Accuracy \\
    \midrule
    Majority class (baseline) & 0.355 & 0.551 \\
    Sensor-only  & 0.756 $\pm$ 0.011 & 0.775 $\pm$ 0.012 \\
    Audio-only   & 0.759 $\pm$ 0.039 & 0.769 $\pm$ 0.038 \\
    Decision PoE & 0.753 $\pm$ 0.013 & 0.772 $\pm$ 0.016 \\
    \textbf{Multimodal (ours)} & 0.781 $\pm$ 0.039 & 0.782 $\pm$ 0.038 \\
    \bottomrule
  \end{tabular}
\end{table}


\subsection{E2: Highlight Validity Against Expert Judgments}
\label{subsec:eval-alignment}

\begin{figure*}[t]
  \centering
  \includegraphics[width=0.92\linewidth]{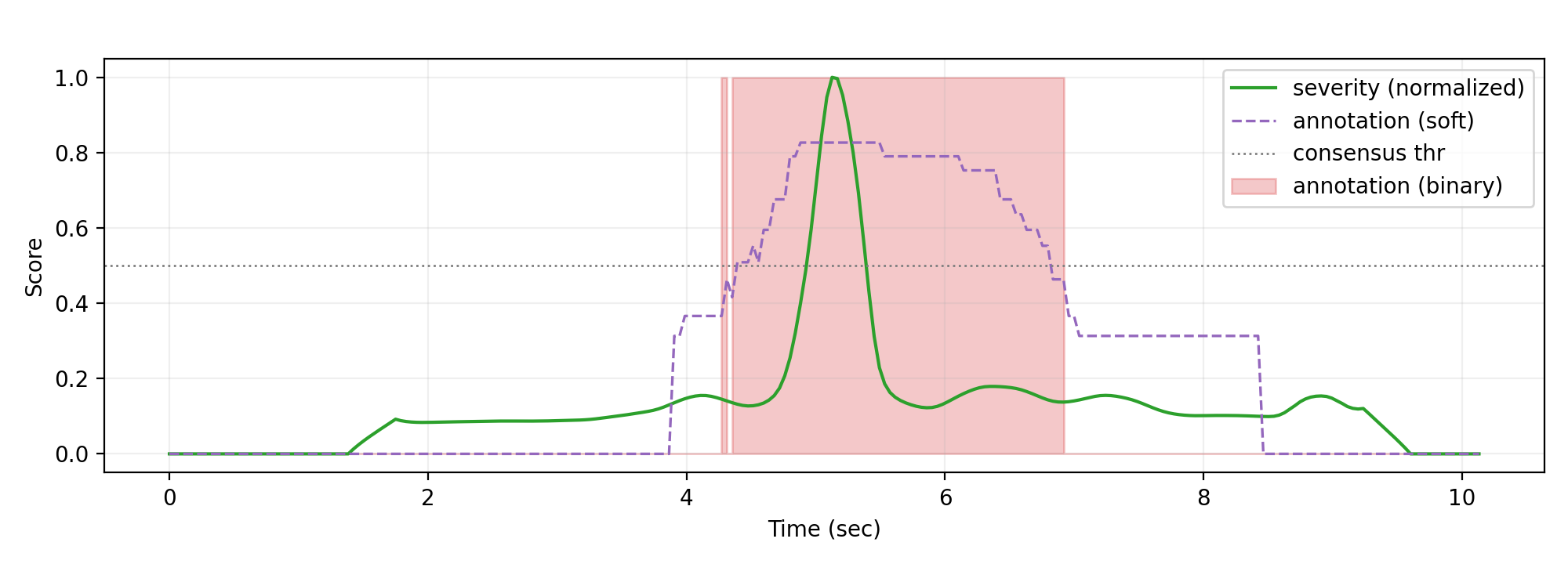}
  \caption{\textbf{Example alignment between model highlight score and expert consensus for an amateur-labeled clip.} The green curve shows the model's normalized highlight score, the dashed curve shows the soft consensus signal from expert spans, the dotted line indicates the consensus threshold, and the shaded region marks the thresholded consensus mask.}
  \Description{Time-series plot showing a green highlight curve, a dashed purple consensus curve, a dotted threshold line, and a red shaded region indicating the binarized consensus span.}
  \label{fig:consensus_overlay}
\end{figure*}

E2 evaluates whether the model-derived highlight signal concentrates on review spans independently marked by human experts.
Figure~\ref{fig:consensus_overlay} shows an example alignment between the model's highlight curve and the expert consensus signal.
We compute agreement against the soft consensus curve using Pearson correlation and against the thresholded consensus mask using Average Precision and ROC-AUC.
Metrics are computed per clip and averaged over the 20 annotated clips.
We include the random-ranking baseline for reference.
For each annotated clip, we use predictions from the cross-validation fold in which the performer is held out (performer-disjoint, out-of-fold).
Table~\ref{tab:alignment_metrics} summarizes these agreement results: the multimodal model obtains the highest Pearson correlation, AP, and ROC-AUC, with AP and ROC-AUC reflecting improved ranked and thresholded alignment with expert-marked spans.
Specifically, the multimodal model reaches Pearson $r{=}0.612$, AP $0.567$, and ROC-AUC $0.753$, outperforming the random-ranking baseline and modestly improving over the strongest unimodal baseline on AP and ROC-AUC.
The small Pearson difference between the sensor-only and multimodal models suggests that multimodal fusion mainly improves prioritization of review spans rather than substantially changing the overall shape of the highlight curve.

\begin{table}[t]
  \centering
  \caption{Agreement between model highlight score and expert consensus on the randomly sampled 20-clip set. Higher is better.}
  \Description{Agreement table for the 20-clip annotation study. Random ranking gives Pearson about 0, average precision 0.20, and ROC-AUC 0.50. Audio-only scores 0.590, 0.539, and 0.720. Sensor-only scores 0.606, 0.546, and 0.732. The multimodal model is highest on all three metrics with Pearson 0.612, average precision 0.567, and ROC-AUC 0.753.}
  \label{tab:alignment_metrics}
  \begingroup\small
  \begin{tabular}{@{}lccc@{}}
    \toprule
    Model & Pearson ($\uparrow$) & AP ($\uparrow$) & ROC-AUC ($\uparrow$) \\
    \midrule
    Random ranking baseline & $\approx 0$ & 0.20 & 0.50 \\
    Audio-only   & 0.590 & 0.539 & 0.720 \\
    Sensor-only  & 0.606 & 0.546 & 0.732 \\
    \textbf{Multimodal (ours)} & 0.612 & 0.567 & 0.753 \\
    \bottomrule
  \end{tabular}
  \endgroup
\end{table}

The alignment scores show that the highlights produced from take-level supervision often overlap with expert-agreed review spans.
This result is consistent with the intended interaction of presenting time-aligned spans for replay and review.
Crucially, the time-localized signal derived from temporal attention and evidence retains measurable alignment with expert judgments under the weakly supervised training setup.

\paragraph{Which expert comments are more often captured.}
To understand which kinds of expert comments are more often captured by the highlight score, we analyzed the free-text notes attached to annotated spans.
Across 21 experts, we collected 407 notes.
We categorized notes using a rule-based keyword coding scheme into timing, legato or connection, unevenness, tone or balance, fingering or coordination, and phrasing or ending.
We thresholded each highlight curve to 20\% coverage and counted a note as captured when at least 30\% of its span overlapped the highlight.
Table~\ref{tab:note_alignment} reports capture rates by modality.

Timing and unevenness comments frequently correspond to localized instabilities that experts agree on, and these categories tend to be captured more often.
More nuanced tone and phrasing comments are captured less consistently, matching the intuition that they can be harder to localize from brief technical studies and short clips.

\begin{table}[!t]
  \centering
  \caption{Post-hoc capture rates by comment category and modality. Notes are keyword-coded; a note can contribute to multiple categories. Capture requires at least 30\% overlap between the note span and the highlight mask.}
  \Description{Category-by-category capture-rate table for expert comment spans. Rows are timing, unevenness, legato or connection, tone or balance, fingering or coordination, and phrasing or ending. Across categories, the multimodal model generally matches or exceeds the unimodal models, with the strongest rate at unevenness, 63 percent, and the lowest at phrasing or ending, 30 percent. The final column reports the number of notes in each category.}
  \label{tab:note_alignment}
  \begingroup
  \small
  \setlength{\tabcolsep}{3pt}
  \begin{tabularx}{\columnwidth}{@{}>{\raggedright\arraybackslash}Xcccc@{}}
    \toprule
    Category & \shortstack{Audio\\(\%)} & \shortstack{Sensor\\(\%)} & \shortstack{\textbf{Multimodal}\\\textbf{(ours, \%)}} & $n$ \\
    \midrule
    Timing                  & 46 & 57 & 53 & 99 \\
    Unevenness              & 56 & 60 & 63 & 36 \\
    Legato/connection       & 39 & 41 & 42 & 90 \\
    Tone/balance            & 45 & 46 & 46 & 114 \\
    Fingering/coordination  & 40 & 41 & 44 & 99 \\
    Phrasing/ending         & 25 & 20 & 30 & 20 \\
    \bottomrule
  \end{tabularx}
  \endgroup
\end{table}

\subsection{E3: Highlight Robustness Under Input Corruption}
\label{subsec:eval-fusion}

Interactive use depends not only on whether a model can classify expertise (E1), but also on whether its time-aligned highlights remain stable enough for repeated review of the same take.
We therefore stress-test visualization robustness by corrupting one modality at inference time without retraining and measuring how much the displayed highlight signal changes. Here E3 focuses on corruption robustness within a take, complementing future longitudinal analyses of how localized review targets evolve across repeated practice takes.

\paragraph{Perturbations.}
We inject corruption at inference time.
For audio, we add Gaussian noise at target SNRs of 20/10/5/0\,dB.
For sensors, we randomly drop sensor frames at rates 10/30/50\% (dropped frames are zero-filled to preserve the fixed $T{=}1000$ time base).
For audio corruption, we recompute the acoustic quality indicators $\mathbf{q}$ on the corrupted signal to mirror deployment-time behavior of the recording-quality-aware gate.
For all perturbations, we rerun the same alignment, masking, and interpolation pipeline used at test time before applying the unimodal or multimodal encoder. The corruption is therefore introduced before the shared preprocessing stack, and the single-encoder controls inherit the same alignment and interpolation updates as the multimodal model.

\noindent Stability metrics for time-aligned highlights.
Let $\tilde{s}^{\mathrm{clean}}_{1:T}$ be the per-clip normalized highlight score used by the UI under clean inputs, and $\tilde{s}^{\mathrm{corr}}_{1:T}$ the corresponding highlight score under corruption.
We quantify visualization degradation with two complementary measures:
(1) Highlight-curve stability, the Pearson correlation $r$ between $\tilde{s}^{\mathrm{clean}}$ and $\tilde{s}^{\mathrm{corr}}$ (computed over non-silent frames), and
(2) Highlight-mask stability, the intersection-over-union (IoU) between thresholded highlight masks obtained by selecting the top 20\% frames by $\tilde{s}$ (a density comparable to our E2 operating point).
For the multimodal model, we additionally report the sensor share $w_S$ (sum of gate weights assigned to sensor-side candidates), which summarizes the post-softmax fusion weights under each corruption condition.

Table~\ref{tab:robustness} summarizes the observed stability pattern.
Metrics are computed on held-out test clips within each fold and reported as mean $\pm$ SD over folds.

\paragraph{Interpretation.}
Audio corruption strongly affects audio-only highlights and, because the gate is conditioned on acoustic quality indicators, shifts the multimodal gate toward the sensor stream (higher $w_S$).
Sensor dropout primarily affects sensor-side representations.
Since the gate does not receive an explicit sensor-failure indicator, changes in $w_S$ reflect the fused representation after preprocessing rather than direct sensor-failure detection.
Overall, the multimodal highlights remain more stable than the directly corrupted unimodal baseline across these perturbations.

\begin{table*}[t]
  \centering
  \caption{\textbf{Visualization robustness pattern.} Each cell reports highlight stability relative to the clean condition: Pearson $r$ between normalized highlight curves ($\tilde{s}$) and IoU between thresholded highlight masks (top 20\% frames). Audio-only and sensor-only are single-encoder controls under the same shared alignment, masking, and interpolation preprocessing. For the multimodal model we additionally report mean sensor share of fusion weights ($w_S$).}
  \Description{Robustness table comparing audio-only, sensor-only under shared preprocessing, and multimodal highlights under clean input, four levels of audio noise, and three levels of sensor dropout. Audio corruption strongly hurts audio-only stability, while the multimodal model degrades less and increases the reported sensor share. Sensor dropout hurts sensor-only more than multimodal, and the reported multimodal sensor share decreases as dropout increases.}
  \label{tab:robustness}
  {\small
  \setlength{\tabcolsep}{6pt}
  \renewcommand{\arraystretch}{1.04}
  \begin{tabular}{@{}lccc@{}}
    \toprule
    Condition & Audio-only & \shortstack{Sensor-only\\(shared prep.)} & \textbf{Multimodal (ours)} \\
    \midrule
    clean
      & \begin{tabular}[c]{@{}c@{}}$r=1.00$\\IoU$=1.00$\end{tabular}
      & \begin{tabular}[c]{@{}c@{}}$r=1.00$\\IoU$=1.00$\end{tabular}
      & \begin{tabular}[c]{@{}c@{}}$r=1.00$\\IoU$=1.00$\\$w_S=0.55\pm0.10$\end{tabular} \\
    \midrule
    audio SNR 20\,dB
      & \begin{tabular}[c]{@{}c@{}}$r=0.82\pm0.08$\\IoU$=0.42\pm0.08$\end{tabular}
      & \begin{tabular}[c]{@{}c@{}}$r=0.99\pm0.01$\\IoU$=0.97\pm0.02$\end{tabular}
      & \begin{tabular}[c]{@{}c@{}}$r=0.90\pm0.06$\\IoU$=0.55\pm0.10$\\$w_S=0.62\pm0.10$\end{tabular} \\
    audio SNR 10\,dB
      & \begin{tabular}[c]{@{}c@{}}$r=0.65\pm0.10$\\IoU$=0.30\pm0.07$\end{tabular}
      & \begin{tabular}[c]{@{}c@{}}$r=0.97\pm0.03$\\IoU$=0.90\pm0.05$\end{tabular}
      & \begin{tabular}[c]{@{}c@{}}$r=0.82\pm0.08$\\IoU$=0.46\pm0.10$\\$w_S=0.72\pm0.10$\end{tabular} \\
    audio SNR 5\,dB
      & \begin{tabular}[c]{@{}c@{}}$r=0.47\pm0.12$\\IoU$=0.22\pm0.06$\end{tabular}
      & \begin{tabular}[c]{@{}c@{}}$r=0.93\pm0.05$\\IoU$=0.80\pm0.08$\end{tabular}
      & \begin{tabular}[c]{@{}c@{}}$r=0.74\pm0.10$\\IoU$=0.38\pm0.10$\\$w_S=0.81\pm0.09$\end{tabular} \\
    audio SNR 0\,dB
      & \begin{tabular}[c]{@{}c@{}}$r=0.30\pm0.14$\\IoU$=0.15\pm0.05$\end{tabular}
      & \begin{tabular}[c]{@{}c@{}}$r=0.86\pm0.08$\\IoU$=0.66\pm0.10$\end{tabular}
      & \begin{tabular}[c]{@{}c@{}}$r=0.62\pm0.12$\\IoU$=0.29\pm0.10$\\$w_S=0.87\pm0.08$\end{tabular} \\
    \midrule
    sensor dropout 10\%
      & \begin{tabular}[c]{@{}c@{}}$r=0.99\pm0.01$\\IoU$=0.97\pm0.02$\end{tabular}
      & \begin{tabular}[c]{@{}c@{}}$r=0.96\pm0.02$\\IoU$=0.92\pm0.03$\end{tabular}
      & \begin{tabular}[c]{@{}c@{}}$r=0.98\pm0.01$\\IoU$=0.94\pm0.02$\\$w_S=0.52\pm0.08$\end{tabular} \\
    sensor dropout 30\%
      & \begin{tabular}[c]{@{}c@{}}$r=0.99\pm0.01$\\IoU$=0.96\pm0.02$\end{tabular}
      & \begin{tabular}[c]{@{}c@{}}$r=0.88\pm0.05$\\IoU$=0.78\pm0.06$\end{tabular}
      & \begin{tabular}[c]{@{}c@{}}$r=0.96\pm0.02$\\IoU$=0.88\pm0.04$\\$w_S=0.40\pm0.10$\end{tabular} \\
    sensor dropout 50\%
      & \begin{tabular}[c]{@{}c@{}}$r=0.99\pm0.01$\\IoU$=0.95\pm0.03$\end{tabular}
      & \begin{tabular}[c]{@{}c@{}}$r=0.75\pm0.08$\\IoU$=0.55\pm0.08$\end{tabular}
      & \begin{tabular}[c]{@{}c@{}}$r=0.94\pm0.03$\\IoU$=0.82\pm0.06$\\$w_S=0.25\pm0.12$\end{tabular} \\
    \bottomrule
  \end{tabular}
  }
\end{table*}


\section{Discussion}

\subsection{Evaluation Implications for Review Cues}
Profy's main design goal is to narrow the search space of solo practice by turning a complete take into locally reviewable spans.
The evaluation supports this goal in the setting of amateur-labeled short technique studies: the model separates expert-labeled and amateur-labeled takes under performer-disjoint splits, and its displayed highlights align with expert-marked review spans on the 20-clip annotation set.
These findings support Profy as a triage tool for focused listening and replay, while longitudinal effects on learning, retention, and teacher--student communication remain open questions.

\paragraph{Why Multimodality Matters for Localized Interaction.}
Because Pearson differs only slightly between the sensor-only and multimodal models in the 20-clip annotation set (0.606 vs.\ 0.612; Table~\ref{tab:alignment_metrics}), we interpret multimodality's main benefit here as improved ranked prioritization and robustness rather than as a large shift in the overall highlight curve.

First, multimodal integration provides modest gains in thresholded discrimination while Pearson remains close to the strongest unimodal baseline. Practice interfaces typically operate on thresholded or ranked highlights, and the multimodal model improves AP and ROC-AUC relative to sensor-only (AP $0.546$ to $0.567$; ROC-AUC $0.732$ to $0.753$; Table~\ref{tab:alignment_metrics}). This suggests that multimodal cues may help prioritize review spans, even though the overall highlight curve remains similar under clean conditions.

Second, multimodal inputs offer stability in conditions where audio-only capture is compromised.
Real-world practice environments often suffer from background noise, inconsistent microphone placement, silent gaps, or reverberation.
Our stress tests in E3 demonstrate that while audio-only performance degrades sharply under corruption, the multimodal model maintains greater stability by leveraging sensor data (Table~\ref{tab:robustness}).
This robustness is particularly relevant for usage scenarios involving immediate review in uncontrolled environments.

Beyond robustness, sensors likely capture micro-features that are pedagogically salient but acoustically subtle, such as release profiles, key-travel microstructure, and minute timing offsets masked by room acoustics.
Our pipeline conditions reliability-aware gating on acoustic quality indicators, effectively creating a mechanism to weigh sensor data more heavily when the audio channel is unreliable.
Future work should quantify this relationship more directly, for instance by stratifying agreement metrics by audio-quality bins (e.g., low signal-to-noise ratio subsets) to verify whether multimodal gains are concentrated in regimes where audio-only analysis becomes less reliable.

\subsection{Pedagogical Interpretation of Highlighted Spans}
The current single-channel highlight is designed as a low-friction entry point for practice triage: it helps learners decide where to listen next before moving to finer-grained interpretation.
To relate quantitative alignment metrics to pedagogical practice, we analyzed free-text comments attached to expert-marked spans.
The exploratory keyword-based coding of 407 notes from 21 experts reveals distinct patterns in what the model captures (Table~\ref{tab:note_alignment}).

The system appears most reliable when addressing issues that are short and time-specific.
Comments regarding timing and unevenness often refer to compact moments---such as a brief rush or a localized inter-hand drift---that map naturally to a highlight-and-loop workflow.
Conversely, issues related to legato, tone, and phrasing are frequently expressed over longer, less distinct spans or as global impressions.
These characteristics are intrinsically difficult to represent with sparse, time-localized highlighting.
This distinction implies that highlight visualization should be framed as a tool for micro-targeting practice spans rather than as an exhaustive critique of musicality.
A natural next step is to retain this single-channel entry layer while coordinating it with separate timing-, connection-, and balance-related evidence channels when such distinctions can be estimated reliably.

Qualitative evidence from annotator notes illustrates this distinction.
Localized issues are typically described in concise terms, such as ``Tempo wavers and speeds up'' or ``Left-right timing drift is noticeable.''
In contrast, global critiques tend to be phrased holistically, such as ``Overall, legato is insufficient, and the natural rise/fall in dynamics is missing.''

Because capture depends on display density, we report results at an operating point intended to reflect plausible practice use.
We thresholded highlights to 20\% coverage and defined capture as a 30\% overlap with the expert span.
These parameters approximate a plausible density for practice review.
While different thresholds would alter absolute capture rates, the qualitative conclusion remains robust: localized issues are compatible with sparse highlighting, whereas long-horizon issues likely require complementary representations, such as longer-window summaries or teacher-authored tags.

\subsection{Diagnosability Across the Task Suite}
\label{subsec:disc_piecewise}

Practical deployment requires an understanding of which technical tasks the model handles most effectively.
We analyzed clip-level accuracy by piece for the multimodal model to determine which exercises are most diagnostic.
Table~\ref{tab:piece_accuracy} summarizes the most and least separable exercises, averaged per-piece across folds.

\begin{table}[!t]
  \centering
  \caption{Top and bottom pieces by mean accuracy for the multimodal model, averaged over 3 folds. Values include mean accuracy and mean Macro-F1. $N$ indicates total support across folds.}
  \Description{Piece-wise ranking table for the multimodal model. The top five rows are scale exercises, led by Scale B Major with accuracy 0.881 and Macro-F1 0.874. The bottom five include both scales and arpeggios, with Scale F-sharp minor lowest at accuracy 0.607 and Macro-F1 0.605. Each row also reports total support, which ranges from 71 to 73 clips.}
  \label{tab:piece_accuracy}
  \begingroup\small
  \begin{tabular}{@{}lccc@{}}
    \toprule
    Piece & Accuracy & Macro-F1 & $N$ \\
    \midrule
    \multicolumn{4}{l}{Top 5} \\
    Scale B Major         & 0.881 & 0.874 & 72 \\
    Scale D$\flat$ Major  & 0.831 & 0.827 & 73 \\
    Scale B$\flat$ minor  & 0.826 & 0.814 & 72 \\
    Scale C Major         & 0.803 & 0.798 & 73 \\
    Scale C$\sharp$ minor & 0.791 & 0.785 & 71 \\
    \midrule
    \multicolumn{4}{l}{Bottom 5} \\
    Scale E$\flat$ minor    & 0.663 & 0.662 & 71 \\
    Arpeggio B$\flat$ minor & 0.660 & 0.655 & 73 \\
    Arpeggio D Major        & 0.648 & 0.644 & 73 \\
    Scale G$\flat$ Major    & 0.635 & 0.623 & 73 \\
    Scale F$\sharp$ minor   & 0.607 & 0.605 & 71 \\
    \bottomrule
  \end{tabular}
  \endgroup
\end{table}

The results indicate that some scalar exercises are highly separable (e.g., Scale B Major, Accuracy = 0.881, Macro-F1 = 0.874), while certain arpeggios and difficult scale conditions are less separable.
This suggests that the ``evenness'' features captured by our sensor and audio encoders may be especially discriminative for some continuous scalar runs, while arpeggios and more difficult scale conditions may require complementary cues.
For interface design, this finding implies that systems supporting arpeggio-heavy repertoire or less separable scale conditions may require higher confidence thresholds or complementary heuristics.

\subsection{Design Implications for Expert Feedback Collection}
\label{subsec:disc_annotation}

Since Profy generates time-localized evidence, the validity of our evaluation and future human-in-the-loop refinements depends on the ability of experts to externalize judgments as editable time spans.
Our formative pilot revealed that interaction friction in the annotation interface can systematically bias these expressions.
Specifically, waveform-only navigation increases the cost of locating short target moments, potentially leading annotators to mark coarser spans rather than fine-grained specific issues.

Annotation interfaces for skill feedback must preserve expert granularity by lowering the cost of finding and describing micro-events while mitigating procedural confounds.
Our experience suggests several specific requirements.
First, interfaces should provide domain-native anchors, such as compact score cues, to serve as semantic landmarks.
Second, controls for micro-event listening---including scrubbing and slow playback---are essential for judging timing and coordination without repeated trial-and-error.
Third, the ability to attach pedagogical rationales allows for the capture of concepts that span boundaries alone cannot convey.
Finally, the protocol design itself plays a role; randomizing item order and implementing completion checks can reduce fatigue and prevent data omission.

A central premise of this work is the reduction of supervision costs.
We avoid the most expensive form of supervision---dense, time-localized technique labels---by relying on take-level supervision.
While cohort collection remains important for discriminative learning, future systems may explore lighter-weight validation pipelines, such as combining self-reported background data with lightweight verification or using active learning to query experts only on uncertain cases.

The same design logic may also apply beyond piano when the task provides a segmentable time series or procedural sequence, domain-native anchors for localization, and coarse labels that correlate with quality. In medical training, these anchors may be procedural steps and spatially situated simulation cues; in vocational training, they may be machine-task stages and learner-following status in AR tutoring; and in craft practice, they may be material states, tool--body coordination, or editable teaching traces that help surface tacit knowledge~\cite{Xu2021HMDMedicalReview,Huang2021AdapTutAR,Chen2025OrigamiSensei,Ji2025ReshapingCraftLearning,Hao2025TacitTeaching}. Under these conditions, a system can generate reviewable local targets for user action without dense ground-truth annotation, and the relevant anchors can differ from the score- and timeline-based cues used in piano practice.

\subsection{Implications for Practice Support Interfaces}
\begin{figure}[t]
  \centering
  \includegraphics[width=\linewidth]{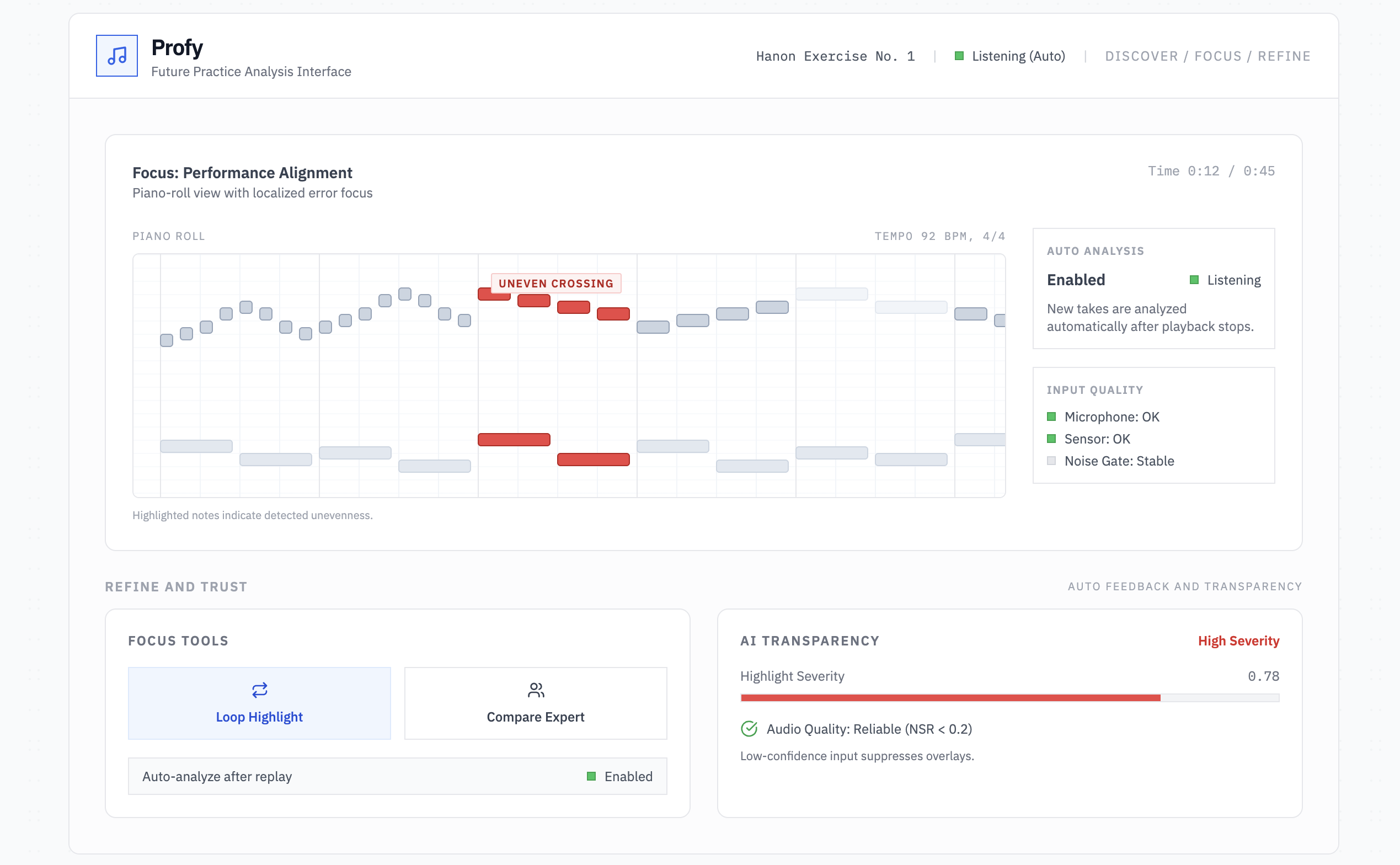}
  \caption{Conceptual practice interface sketch showing timeline highlights, loop controls, exemplar comparison, and input-quality cues aligned with the Discover--Focus--Refine workflow.}
  \Description{Mock interface with a piano-roll view, red highlight overlays, loop and compare buttons, and panels indicating microphone and sensor status.}
  \label{fig:future_ui}
\end{figure}

The interface presentation shapes how the model output is interpreted, because the same signal can be constructive or discouraging depending on the interaction design.
Figure~\ref{fig:future_ui} illustrates a conceptual interface aligned with a Discover--Focus--Refine workflow.

The interface should treat highlights as actionable suggestions rather than definitive judgments.
In the Discover phase, the system presents a limited number of time-aligned spans.
In the Focus phase, users convert these spans into practice units using looping and speed controls.
Finally, the Refine phase supports iteration by emphasizing changes over time, rather than presenting a fixed judgment map.
In a typical session, a learner may first inspect a small set of highlighted spans, audition one passage with looping or slow playback, make a focused adjustment, and then compare a revised take against the previous one using the same localized view.

Although our evaluation used a fixed threshold, practical applications should treat highlight density as a user-controlled parameter.
Exposing a density control allows users to balance breadth against focus.
A conservative default density (e.g., 10--15\%) may help prevent discouragement, and reliability-aware suppression can further reduce highlight density when input reliability is low.
Presets could also be adapted to skill level, with lower density for beginners and higher density for advanced micro-refinement.

At present, the system is optimized to localize where to inspect next; identifying what to correct can then be supported by replay, listening, experimentation, or teacher input.
In unguided practice, learners often face a broad search space when deciding which part of a take deserves attention.
By narrowing that search space to specific time spans, Profy can reduce the cost of self-guided review even when the correction strategy still depends on listening, experimentation, or teacher input.

\subsection{Limitations and Responsible Use}
Three aspects define the scope and responsible design of our proposed approach.

First, this paper validates the displayed cue against expert span judgments rather than longitudinal learning outcomes. The present evidence supports its use as a local review aid for short technique studies. Outcomes such as retention, transfer, motivation, and teacher--student communication require separate intervention studies.

Second, the evidence signals summarize model-relevant local evidence under a clip-level objective. In practice, the cues are designed to function as pointers for replay and listening, helping users narrow attention before forming a musical judgment, since highlighted passages may reflect expressive variation, tempo differences, recording conditions, or sensing artifacts in addition to pedagogically actionable technique. This behavior follows naturally from weak supervision, where any stable predictor of the coarse label can contribute to the decision. Accordingly, peak saliency serves most productively as a prompt for hypothesis-driven listening rather than as a standalone diagnosis.

Third, the definition of ``expertise'' is implicit in the training data and therefore reflects the stylistic norms represented in the cohort, repertoire, and recording protocol. While such norms often align with mainstream pedagogy, they are not universally appropriate across genres, schools of technique, or individual constraints (e.g., interpretive intentions, physical characteristics, or instrument/environment differences). For this reason, the system is most useful as a reflective cue that complements musical judgment rather than replacing it with a single normative grade, especially outside the regime represented by our corpus (e.g., longer expressive repertoire, different genres, or different recording setups).

To mitigate these risks, we recommend specific interface countermeasures aligned with the limitations above.
To address potential false positives and normative interpretation (e.g., rubato incorrectly presented as a flaw), highlights should be presented as neutral ``difference cues'' rather than faults, accompanied by uncertainty framing and optional comparative exemplars (ideally multiple references or teacher-selected examples) so users can judge whether a difference is intentional.
To address false negatives and long-horizon issues that sparse, local cues cannot cover (e.g., phrasing or global legato), the interface should include long-window summaries and segment comparison views that encourage holistic listening and phrase-level review.
To prevent over-reliance---where learners passively follow highlights as prescriptions---the workflow should distinguish between ``Explore'' and ``Verify'' modes, requiring active listening (and optionally a brief self-explanation) before committing a span as a practice target.
Finally, to avoid demotivation and to support constructive learning trajectories, the system should emphasize improvement deltas across sessions, visualize achievements alongside remaining practice targets, and allow users to tune highlight density (or suppress cues) when they are not helpful.


\section{Conclusion}
We introduced Profy, a weakly supervised system that identifies time-aligned practice targets using only take-level expertise labels.
Evaluation on 1{,}083 valid takes from 73 pianists used for modeling and evaluation shows that our localized highlights align meaningfully with expert consensus (Pearson $r{\approx}0.61$), even without training on granular annotations.
By producing inspectable time-localized cues rather than only global assessment, Profy shows how weak supervision can support scrubbing and looping while reducing the burden of dense annotation.
This work suggests a broader design pattern for interactive support in piano and adjacent embodied skill domains: transforming coarse quality labels into reviewable local spans when the task provides repeated temporal units or procedural steps, replayable actions, and domain-native anchors for inspection.


\begin{acks}
This work was supported by the JST Moonshot R\&D Program Grant Number JPMJMS2012, JST ASPIRE Grant Number JPMJAP2503, and JST CRONOS Grant Number JPMJCS24N8.
\end{acks}

\bibliographystyle{ACM-Reference-Format}
\bibliography{references/references}

\end{document}